\documentclass[twocolumn,trackchanges]{aastex631}
\usepackage[T1]{fontenc}
\usepackage{fontawesome}
\usepackage{color}
\usepackage{amsmath}
\usepackage{natbib}
\usepackage{ctable}
\usepackage{bm}
\usepackage[normalem]{ulem} 
\usepackage{xspace}
\usepackage{paralist}
\usepackage{fontawesome}
\usepackage[caption=false]{subfig}

\newcommand{\simbig}{{\sc SimBIG}}

\defcitealias{simbig_letter}{H22}
\defcitealias{simbig_challenge}{H23b}

\begin{document}

\title{\simbig: Cosmological Constraints using Simulation-Based Inference of Galaxy Clustering with Marked Power Spectra}

\author[0000-0002-0637-8042]{Elena Massara}
\altaffiliation{elena.massara.cosmo@gmail.com}
\affiliation{Waterloo Centre for Astrophysics, University of Waterloo, 200 University Ave W, Waterloo, ON N2L 3G1, Canada}
\affiliation{Department of Physics and Astronomy, University of Waterloo, 200 University Ave W, Waterloo, ON N2L 3G1, Canada}

\author[0000-0003-1197-0902]{ChangHoon Hahn}
\affiliation{Department of Astrophysical Sciences, Princeton University, Princeton NJ 08544, USA} 

\author{Michael Eickenberg}
\affiliation{Center for Computational Mathematics, Flatiron Institute, 162 5\textsuperscript{th} Avenue, New York, NY 10010, USA}

\author{Shirley Ho}
\affiliation{Center for Computational Astrophysics, Flatiron Institute, 162 5\textsuperscript{th} Avenue, New York, NY 10010, USA}

\author{Jiamin Hou}
\affiliation{Department of Astronomy, University of Florida, 211 Bryant Space Science Center, Gainesville, FL 32611, USA}
\affiliation{Max-Planck-Institut f\"ur Extraterrestrische Physik, Postfach 1312, Giessenbachstrasse 1, 85748 Garching bei M\"unchen, Germany}

\author{Pablo Lemos}
\affiliation{Department of Physics, Universit\'{e} de Montr\'{e}al, Montr\'{e}al, 1375 Avenue Th\'{e}r\`{e}se-Lavoie-Roux, QC H2V 0B3, Canada}
\affiliation{Mila - Quebec Artificial Intelligence Institute, Montr\'{e}al, 6666 Rue Saint-Urbain, QC H2S 3H1, Canada}
\affiliation{Center for Computational Astrophysics, Flatiron Institute, 162 5\textsuperscript{th} Avenue, New York, NY 10010, USA}

\author{Chirag Modi}
\affiliation{Center for Computational Mathematics, Flatiron Institute, 162 5\textsuperscript{th} Avenue, New York, NY 10010, USA}
\affiliation{Center for Computational Astrophysics, Flatiron Institute, 162 5\textsuperscript{th} Avenue, New York, NY 10010, USA}

\author[0000-0001-8841-9989]{Azadeh Moradinezhad Dizgah}
\affiliation{Laboratoire d’Annecy de Physique Théorique (CNRS/USMB), F-74940 Annecy, France}
\affiliation{D\'epartement de Physique Th\'eorique, Universit\'e de Gen\`eve, 24 quai Ernest Ansermet, 1211 Gen\`eve 4, Switzerland}

\author{Liam Parker}
\affiliation{Center for Computational Astrophysics, Flatiron Institute, 162 5\textsuperscript{th} Avenue, New York, NY 10010, USA}
\affiliation{Department of Physics, University of California, Berkeley, CA 94720, USA}

\author[0000-0003-0055-0953]{Bruno Régaldo-Saint Blancard}
\affiliation{Center for Computational Mathematics, Flatiron Institute, 162 5\textsuperscript{th} Avenue, New York, NY 10010, USA}


\date{\today}

\begin{abstract}
We present the first $\Lambda$CDM cosmological analysis performed on a galaxy survey using marked power spectra. The marked power spectrum is the two-point function of a marked field, where galaxies are weighted by a function that depends on their local density. The presence of the mark leads these statistics to contain higher-order information of the original galaxy field, making them a good candidate to exploit the non-Gaussian information of a galaxy catalog. In this work we make use of \simbig, a forward modeling framework for galaxy clustering analyses, and perform simulation-based inference using normalizing flows to infer the posterior distribution of the $\Lambda$CDM cosmological parameters. We consider different mark configurations (ways to weight the galaxy field) and deploy them in the \simbig~pipeline to analyze the corresponding marked power spectra measured from a subset of the BOSS galaxy sample. We analyze the redshift-space mark power spectra decomposed in $\ell = 0, 2, 4$ multipoles and include scales up to the non-linear regime. Among the various mark configurations considered, the ones that give the most stringent cosmological constraints produce posterior median and $68\%$ confidence limits on the growth of structure parameters equal to $\Omega_m=0.273^{+0.040}_{-0.030}$ and $\sigma_8=0.777^{+0.077}_{-0.071}$. Compared to a perturbation theory analysis using the power spectrum of the same dataset, the \simbig~marked power spectra constraints on $\sigma_8$ are up to $1.2\times$ tighter, while no improvement is seen for the other cosmological parameters. 

\end{abstract}


\section{Introduction} \label{sec:intro}

The statistical properties of the three-dimensional (3D) spatial distribution of galaxies can inform us about the content and evolution of the Universe. In the past years, the  Baryon Oscillation Spectroscopic Survey ~\citep[BOSS,][]{Eisenstein2011, Dawson2013} observed over a million galaxies in a volume of few Gpc$^3$ down to redshift $z\sim 0.7$. Current and upcoming spectroscopic surveys, such as DESI~\citep{desi1,desi2}, Euclid~\citep{Laureijs2011}, PFS~\citep{Takada2014, Tamura2016}, and the Roman Space Telescope~\citep{Spergel2015, Wang2022}, will observe much larger and denser galaxy samples up to redshift $z\sim 3$, providing us with much richer datasets exhibiting information in new unexplored regimes. On the one hand, large volumes and high redshift measurements will allow us to study the properties of galaxy clustering up to larger scales and earlier times. On the other end, higher galaxy density catalogs will contain information about the small-scale clustering and they will include faint galaxies that populate cosmic voids, allowing us to probe the very low-density regions of the Universe.

The standard way to perform the cosmological analysis with galaxy surveys is to describe the 3D galaxy distribution via a summary statistic, model it via a theoretical framework with free parameters encoding the cosmological information, and use Bayesian inference to estimate the posterior distribution of the cosmological parameters. Until recently, most analyses of galaxy clustering have employed the power spectrum multipoles, $P_\ell(k)$, or their configuration space counterpart, the two-point correlation function. They are usually modeled using the perturbation theory (PT) of large-scale structure~\citep[see][for a review]{Bernardeau2002,Desjacques2016}, which is valid only on linear and weakly nonlinear scales and causes any analysis to be restricted to large scales (typically $k<0.2~h ${\rm Mpc}$^{-1}$). Then, the Bayesian inference on the cosmological parameters of the PT model is performed assuming a Gaussian form for the likelihood. This framework has been very successful in retrieving cosmological information from the large-scale distribution of galaxies in the Sloan Digital Sky Survey (SDSS) and BOSS within it \citep[\emph{e.g.}][]{Beutler2017, damico2020, Ivanov2020, Chen2022}.

Since current and future surveys will observe a denser galaxy population, the need to go to smaller scales and leverage non-Gaussian information has become pressing. Therefore, increasing attention has been drawn to summary statistics beyond the two-point function. 
Multiple alternatives have been proposed, studied in Fisher forecast analyses \citep[e.g.,][]{Hahn2021,Naidoo2022,Eickenberg2022,Valogiannis2022,Kreisch2022,Hou2023Fisher,Paillas2023}, and applied to galaxy surveys \citep[e.g.,][]{DAmico2022,Ivanov2023,simbig_wave2,Paillas2023,Contarini2023,Valogiannis2023}, proving that there is non-Gaussian information on small nonlinear scales. 

In this paper, we consider one of these alternative summary statistics, the marked power spectrum $M_\ell(k)$ \citep[see e.g.,][for a theoretical introduction]{Sheth2005,White2016,Philcox2020}. Marked power spectra are extensions of the power spectrum where galaxies are weighted by a mark that depends on their environmental density. The presence of the mark introduces non-Gaussian information in the statistics, and the choice of up-weighting objects in low-density regions compared to those in high-density environments allows us to retrieve information from voids, usually under-represented in $P_\ell(k)$. Fisher forecast showed that $M_\ell(k)$ contains additional information than $P_\ell(k)$, especially on $\sigma_8$, the neutrino mass scale \citep{Massara2023,Massara2021}, and theories of modified gravity \citep{Valogiannis2018}. While a PT framework to describe marked power spectra has been developed in \cite{Philcox2020} and \cite{Philcox2021}, the need to probe small scales requires going beyond PT. Moreover, the assumption of a Gaussian likelihood usually applied to galaxy clustering analyses does not hold in general and for every summary statistic. 

There are at least two different approaches that attempt to overcome the shortfalls of the standard clustering analysis, and they both use simulation-based models of the 3D galaxy distribution. Simulation-based models allow us to describe any summary statistic up to nonlinear scales, with a scale limit set by the simulator itself. They also enable the possibility of including the survey geometry and observational systematics directly in the forward model. Indeed, while weighting schemes to account for incompleteness, target failure, and fiber collision (for instruments with fiber-fed spectrographs) have been developed for analyses with $P_\ell(k)$ \citep{Guo2012,Hahn2017,Pinol2017,Bianchi2018}, they have not been designed or demonstrated to be effective for other summary statistics. 

The first approach is given by emulators, which are simulation-based models that, like PT, aim at describing the mean of the summary statistics but can be used up to nonlinear scales. However, emulators are usually coupled to the standard Bayesian inference pipeline that assumes a Gaussian likelihood \citep[see e.g.,][]{Kobayashi2022,Zhai2023Aemulus,Paillas2023CMASS}. The second method is represented by Simulation-Based Inference (SBI) \citep[see e.g.,][for a review]{Cranmer2020}, which combines simulation-based modeling and an inference pipeline that does not require the assumption of a Gaussian likelihood: The likelihood is implicitly learned from the forward model. 

Recently, \cite{simbig_letter} and \cite{simbig_challenge}\footnote{Hereafter \citepalias{simbig_letter} and \citepalias{simbig_challenge}} introduced SIMulation-Based Inference of Galaxies (\simbig), the first forward modeling and SBI framework to analyze galaxy clustering. \simbig~provides a forward model to produce realistic galaxy mock catalogs and summary statistics starting from high-fidelity $N$-body simulations at different cosmologies, incorporating systematic effects such as survey geometry and fiber collisions directly at the 3D galaxy field level. Moreover, it uses neural density estimators (NDE) trained to infer the posterior distribution of the cosmological parameters. The \simbig~pipeline has successfully been deployed to perform the cosmological analysis of the Sloan Digital Sky Survey (SDSS)-III Baryon Oscillation Spectroscopic Survey (BOSS) with many different summary statistics: the bispectrum \citep{simbig_bk}, wavelet scattering statistics \citep{simbig_WST}, a Convolutional Neural Networks (CNN) compression of the galaxy field \citep{simbig_cnn}, and skew spectra \citep{simbig_skewspec}. 

In this paper, we extend the \simbig~pipeline to another summary statistic: the marked power spectrum. We perform the first cosmological analysis with marked power spectra using \simbig~applied to the SDSS-III BOSS galaxy catalog and obtain constraints on the 5 cosmological parameters of the $\Lambda$CDM model. The paper is structured as follows: Section~\ref{sec:data} describes the observational data and \simbig~forward model used to generate the synthetic catalogs, Section~\ref{sec:method} explains how marked power spectra are measured in \simbig, the SBI framework and its validation, Sections~\ref{sec:results} and \ref{sec:discussion} present and discuss the results of our analysis, and Sections~\ref{sec:conclusions} presents our conclusions. 

\section{Observational and Synthetic Data}
\label{sec:data}
In this section, we describe the observed dataset used to do the inference and the synthetic data used to train, validate, and test the SBI.

\subsection{Observations: BOSS CMASS SGC Galaxies}

We analyze the Luminous Red Galaxies (CMASS) from the 12th data release of the Baryon Oscillation Spectroscopic Survey (BOSS)~\citep{Eisenstein2011, Dawson2013}. Because of the limited volume of the simulations used to train the NDEs and the need to fit the observational data within it, we limit the observed catalog to a subset of the Southern Galactic Cap (SGC) of the CMASS sample within specified angular boundaries (${\rm Dec} > -6$ deg. and $-25 < {\rm RA} < 28$ deg.) and redshift range ($0.45 < z < 0.6$), as explained in \citepalias{simbig_letter}. In total, the selected galaxy sample spans approximately $3,600$ square degrees and encompasses $109,636$ galaxies, constituting $70\%$ of the SGC footprint and around $10\%$ of the entire BOSS volume (see \citepalias{simbig_letter} for visuals of the catalog).

\subsection{\simbig~Forward Model}
\label{sec:forward}
SBI requires a forward model, i.e., the ability to simulate an observation, ${\bf x}$, starting from a set of parameters ${\bm \theta}$. Once many observations ${\bf x}_i$ are generated at different parameters' values ${\bm \theta}_i$ drawn from a prior distribution $p({\bm \theta})$, a neural density estimator (NDE) can be trained to approximate the posterior distribution $p({\bm \theta}|{\bf x})\propto p({\bf x}|{\bm \theta})p({\bm \theta})$. Thanks to this procedure, the likelihood $p({\bf x}|{\bm \theta})$ is directly learned from the simulated data, without requiring any strong assumption on its functional form.

We use the forward model of 
\simbig~\citepalias{simbig_letter} that is built via the following steps: (1) an $N$-body simulator predicts the matter field at the redshift of interest, (2) a halo finder identifies halos in the matter field, (3) halos are populated with galaxies, (4) survey geometry and fiber collision effects are added to the galaxy mock catalogs. Below we describe the choices taken to implement these steps to generate the training and validation sets. In the following subsection, we will describe a different implementation of these steps to generate the test sets for the mock challenge. 

The $N$-body simulations used for training and validation are part of the {\sc Quijote} suite \citep{quijote}, a large set of simulations run to train machine learning models and perform Fisher analyses. \simbig~uses the high-resolution boxes (1024 cold dark matter particles in $1~ h^{-3}{\rm Gpc}^{3}$) at redshift $z=0.5$ and arranged in a Latin Hypercube (LH) configuration, which allows us to uniformly sample the space of the 5 $\Lambda$CDM cosmological parameters, ${\bm \theta}_{\rm cosmo}=\left\{\Omega_{\rm b},h,n_{\rm s},\Omega_{\rm m},\sigma_8\right\}$, with priors detailed in Table~\ref{tab:prior}.

\begin{table}
   \centering
   \begin{tabular}{|c|ccc|}
       \hline
       $\bm{\theta}_{\rm cosmo}$ & Prior & {\sc Quijote} & {\sc AbacusSummit} \vspace{-0.1cm}\\
         & ({\sc Quijote} LH) & fiducial & \\
       \hline
       $\Omega_{\rm b}$ & $\mathcal{U}(0.03, 0.07)$ & 0.049 & 0.0493\\
       $h$ & $\mathcal{U}(0.5, 0.9)$ & 0.6711 & 0.6736\\
       $n_s$ & $\mathcal{U}(0.8, 1.2)$ & 0.9624 & 0.9649\\
       $\Omega_{\rm m}$ & $\mathcal{U}(0.1, 0.5)$ & 0.3175 & 0.3152 \\
       $\sigma_8$ & $\mathcal{U}(0.6, 1.0)$ & 0.834 & 0.8080\\
       \hline
   \end{tabular}
   \caption{Prior distributions $p(\bm{\theta}_{\rm cosmo})$ of the cosmological parameters $\bm{\theta}_{\rm cosmo}$ in our inference pipeline, and values of $\bm{\theta}_{\rm cosmo}$ in the {\sc Quijote} and {\sc AbacusSummit} simulations used in the \simbig~ mock challenge.}
   \label{tab:prior}
\end{table}

Halos are identified from the dark matter snapshots using the {\sc Rockstar} halo finder \citep{rockstar}, which can accurately locate halos and substructures within them using phase-space information. Then galaxies are assigned to halos using a Halo Occupation Distribution (HOD) model with 9 parameters, ${\bm \theta}_{\rm HOD}$: 5 parameters from the standard \citep{Zheng2007} model with the addition of 4 parameters describing assembly, concentration, and velocity biases to allow for the flexibility that recent works suggest may be necessary to describe galaxy clustering \citep[e.g.,][]{zentner2016,vakili2019,hadzhiyska2021}. Each halo catalog, at a particular cosmology, is populated 20 times using 20 different sets of the HOD parameters. This allows us to generate 40,000 different galaxy mock catalogs out of 2,000 $N$-body simulations, effectively sampling 40,000 times the HOD parameter space\footnote{We sample with twice the number of samples as in \citepalias{simbig_letter} to have a better representation of different halo-galaxy connections.}.

Finally, the \simbig~forward model applies survey realism. The simulation box is cut into a cuboid and trimmed to reproduce the geometry of our BOSS CMASS SGC subsample, including masking for bright stars, centerpost, bad field, and collision priority. Moreover, fiber collision is applied by removing one galaxy in $60\%$ of pairs of galaxies with angular distance smaller than 62". 

The target parameters of the inference are 14 in total, ${\bm \theta} = \left\{{\bm \theta}_{\rm cosmo},{\bm \theta}_{\rm HOD}  \right\}$. We refer the reader to \citepalias{simbig_challenge} and \citepalias{simbig_letter} for further details on the HOD parameters, the \simbig~forward model in general, and visuals of the final galaxy mock samples.

\subsection{\simbig~Mock Challenge Catalogs}
\label{sec:mock_challenge_data}
To test the robustness of the cosmological inference pipeline, \citepalias{simbig_challenge} introduced the \simbig~mock challenge. This consists of applying the \simbig~ inference pipeline described in Section~\ref{sec:sbi} to the following data sets.

{$\mathtt{TEST0}$}: This test set is generated using the same forward model employed for the training and validation sets (see Section~\ref{sec:forward}), but it exploits a tighter prior for the HOD parameters and a different set of $N$-body simulations obtained from the same simulator. Thus, it serves to test the ability to generalize to unseen data. It is built upon 100 {\sc Quijote} high-resolution $N$-body simulations generated at a fixed cosmology, called the fiducial cosmology of the {\sc Quijote} suite (see Table~\ref{tab:prior} for details). The {\sc Rockstar} halo finder is used to identify halos and the 9-parameter HOD framework described in Section~\ref{sec:forward} is implemented 5 times per $N$-body simulation to obtain a total of 500 realistic galaxy mock catalogs.

$\mathtt{TEST1}$: This test set uses the same $N$-body solver as the training set, but a different halo finder and HOD framework. It is built using the 100 {\sc Quijote} simulations at fiducial cosmology, while halos are identified using the Friend-of-Friend (FoF) algorithm \citep{Davis1985} with linking length parameter set to $0.2$. Galaxies are assigned to halos using the \cite{Zheng2007} 5-parameter HOD model. 
Then CMASS SGC survey realism is applied to the simulations. 
Five different galaxy catalogs are generated for each $N$-body simulation, resulting in a total of 500 mocks. These catalogs are ``out-of-distribution" since they exhibit a halo definition and halo-galaxy connection scheme different from the training set. This test set is used to check the robustness of the inference pipeline against changes in the way astrophysical processes are described and parameterized. 

$\mathtt{TEST2}$: This test set employs a $N$-body solver, a halo finder, and a HOD parametrization that are different from the training set. 200 boxes of $1 h^{-3}{\rm Gpc}^3$ volume are built from 25 simulations in the ``base" configuration of the {\sc AbacusSummit} suite \citep{abacus}, the {\sc CompaSO} halo finder \citep{compaso} is used to identify halos from the dark matter field at $z=0.5$, and the \cite{Zheng2007} 5-parameter HOD model populate halos with galaxies. 5 different sets of HOD parameters are then used to build the galaxy catalogs with survey realism for each simulation box, resulting in  a total of 1,000 mocks. This test set allows us to check for robustness against changes in the definition of halos and the way different $N$-body simulators solve the small-scale matter evolution. It is a second ``out-of-distribution" set.

\section{Method}
\label{sec:method}
In this section, we describe the statistics used to summarize the 3D spatial information content in a galaxy catalog---the marked power spectrum---, the method to retrieve this information---SBI---, and the way we validate the final cosmological inference. 

\subsection{Marked Power Spectra}

\begin{figure}
\begin{center}
    \includegraphics[width=0.9\columnwidth]{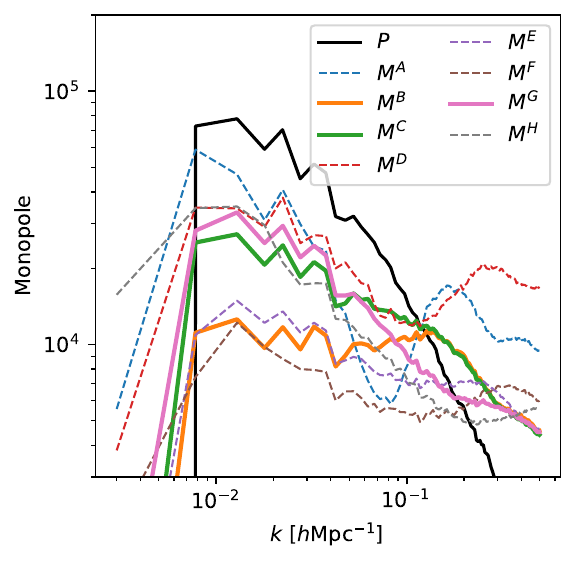}
    \caption{\label{fig:Mk_measured} 
    The monopole of the marked power spectra of our BOSS CMASS SGC subsample for different mark configurations (color-coded).
    We highlight the mark configurations that we select for our analysis ($B$, $C$, and $G$) as solid lines. 
    The black line shows the monopole of the standard power spectrum. 
}
\end{center}
\end{figure}
\begin{table}
\def\arraystretch{1.3}
   \centering
   \begin{tabular}{|c|ccc|}
       \hline
        Name & $R~[h^{-1}{\rm Mpc}]$ & $\qquad p \qquad$ & $\qquad\delta_s\qquad$ \\
       \hline
       $M^A_\ell$ & 30 & 2 & 0.1 \\
       ${\bm M^B_\ell}$ & {\bf 30} & {\bf 1} & {\bf 0.5} \\
       ${\bm M^C_\ell}$ & {\bf 30} & {\bf 0.5} & {\bf 0.1} \\
       $M^D_\ell$ & 20 & 2 & 0.1 \\
       $M^E_\ell$ & 20 & 1 & 0.5 \\
       $M^F_\ell$ & 15 & 1 & 0.1 \\
       ${\bm M^G_\ell}$ & {\bf 15} & {\bf 0.5} & {\bf 0.1} \\
       $M^H_\ell$ & 10 & 1 & 0.1 \\
       \hline
   \end{tabular}
   \caption{\label{tab:Mk_models} List of the mark configurations employed in our pipeline detailing the name of the related marked power spectra and the values for the mark parameters $\rm{R}$, $p$, and $\delta_s$. The models are listed in decreasing values for $\rm{R}$ and $p$. The best models are highlighted in bold and are the ones for which we show and discuss the inferred cosmological constraints. 
   }
\end{table}
The marked power spectrum $M_\ell(k)$ is the Fourier transform of the marked correlation function: the two-point statistic of a marked field, 
\begin{equation}
    \delta_{\rm M}({\bf x}) = \frac{m({\bf x})}{\bar{m}}\left[1+\delta({\bf x})\right]-1,
\end{equation}
obtained by weighting the galaxy field $\delta({\bf x})$ with a mark \citep{White2016}
\begin{equation}
    m({\bf x}) = \left( 1-\frac{\delta_{\rm R}({\bf x})}{1+\delta_s}\right)^{-p}\,,
\end{equation}
where $\delta_{\rm R}({\bf x})$ is the local overdensity around the position ${\bf x}$, i.e. the galaxy overdensity filtered with a top-hat on scale R, $\bar{m}$ is the average of $m({\bf x})$  weighted by the galaxy number density, and R, the bias $\delta_s$, and the exponent $p$ are user-defined parameters. Thus, while only a single power spectrum can be measured from a galaxy field, multiple marked power spectra can be computed by varying the value of its three parameters. The scale $\rm{R}$ defines the maximum distance determining the mark at a given point in space. The exponent $p$ and the bias $\delta_s$ dictate how much the value of the local density $\delta_{\rm R}$ impacts the final mark. When $p$ is close to zero or $\delta_s$ is very large, the mark $m({\bf x})$ tends to unity. Moreover, when $p$ is positive, galaxies in low-density environments are up-weighted compared to those in high-density regions. This means that mark power spectra with $p>0$ are more sensitive to low-density regions, while the opposite happens when $p$ is negative. \cite{Massara2021} and \cite{Massara2023} showed that marked power spectra that up-weight low-density regions carry additional information compared to the power spectrum. In the power spectrum, high-density regions have already and naturally more contribution than low-density ones, since the absolute value of the density contrast of the non-linear galaxy field is larger in peaks than in voids.   

We implement the pipeline to measure marked power spectra in data with survey geometry in the following way. We use a random catalog with many points (>4,000,000) and the same angular and radial selection as the synthetic data and the observations. We estimate $1+\delta_{\rm R}=n_{\rm R}/\bar{n}_{\rm R}$ at each galaxy location using a $k$-d tree algorithm to count the number density of galaxies (numerator) and randoms (denominator) within a distance R. The galaxy number is weighted by systematics and redshift failure weights; the fiber collision weights are not included since their effect is already incorporated in the \simbig~forward model. Then, for a set of mark parameters (${\rm R}, p, \delta_s$), we measure the marked power spectrum moments $M_\ell(k)$ using the {\sc nbodykit} package \citep{nbodykit}. In particular, we use the {\sc ConvolvedFFTPower} function that computes the auto-power spectrum of a field, in our case the marked galaxy field, where each galaxy has been weighted by the mark and the systematics and redshift failure weights. For each synthetic catalog and the observed one, we measure $M_\ell(k)$ with $l=0,2,4$ in bins $\Delta k = 0.005~ h{\rm Mpc}^{-1}$ up the maximum wavelength $k=0.5~h{\rm Mpc}^{-1}$. We concatenate the three multipoles with the galaxy number density $\bar{n}_g$, obtaining a final data vector ${\bf x}=\left\{M_\ell(k),\bar{n}_g\right\}$ with 301 coefficients per galaxy catalog. We do not subtract the shot noise from the monopole of the marked power spectra because its Poissonian estimation \citep{Philcox2020} is an idealized version of the true shot noise and, when subtracted from the measured monopoles, it leads to negative values for specific $k$ bins of some marked power spectra.

We consider different mark configurations (sets of mark parameters) and measure the corresponding marked power spectra in the mock and observed galaxy catalogs. The mark configurations used in this work are listed in Table~\ref{tab:Mk_models} and they have been informed by the Fisher analysis performed in~\cite{Massara2023}. The monopole of the related marked power spectra measured from our subset of the BOSS CMASS SGC sample are displayed in Figure~\ref{fig:Mk_measured} as colored lines, together with the monopole of the standard power spectrum in black. We perform the cosmological analysis for all these marked power spectra, however, we will present the validation and results only for the mark configurations that give robust and interesting cosmological constraints (highlighted in bold in the table and as solid and colored lines in the figure). We will only briefly discuss the other mark configurations (dashed lines in the figure) in Section~\ref{sec:discussion}.

\subsection{Simulation-Based Inference}
\label{sec:sbi}
We use the \simbig~SBI framework to infer the posterior distribution $p({\bm \theta}|{\bf x})$ of the cosmological parameters ${\bm \theta}$ given an observed summary statistic ${\bf x}$. In our case, ${\bf x}$ is a vector containing the galaxy number density and the marked power spectrum multipoles, $M_\ell(k)$. The \simbig~SBI framework uses a neural density estimator (NDE) to estimate the posterior distribution solely from the \simbig~forward model, consisting of 40,000 synthetic samples. Formally, they are 40,000 couples of parameter sets and summary statistic vectors $({\bm \theta}_i,{\bf x}_i)$ drawn from the joint distribution $p({\bm \theta},{\bf x})$. As in previous \simbig~analyses, we use a Masked Autoregressive Flow \citep[MAF;][]{Papamakarios2017} for our NDE from the {\sc sbi} Python package \citep{sbi}.

\begin{table}
    \def\arraystretch{1.2}
   \centering
   \begin{tabular}{|c|c|c|}
       \hline
       Parameter & Range: 1 $M_\ell$ & Range: 2 $M_\ell$\\
       \hline
       N$_{\rm t}$ & [5, 13]& [5, 13] \\
       
       N$_{\rm h}$ & [128, 1024] & [128, 1504]  \\
       
       N$_{\rm b}$ & [2, 5]& [2, 5]  \\
       
       p$_{\rm drop}$ & [0.0, 0.5] & [0.0, 0.5] \\
       
       Batch size&  [2, 100]&  [2, 100] \\
       
       $l_r [\times 10^{-5}]$  & [0.5, 50]& [0.5, 50] \\
       
       \hline
    \end{tabular}
   \caption{List of the hyperparameters optimized using {\sc Optuna} and their sampled ranges when using a single marked power spectrum (central column) and two marked power spectra (right column) as summary statistics. N$_{\rm t}$ is the number of affine transforms in the MAF, N$_{\rm h}$, N$_{\rm b}$, and $p_{\rm drop}$ are the number of hidden units and blocks, and the dropout probability in the MLP, Batch is the batch size, and $l_r$ is the learning rate.}
   \label{tab:hyperparams}
\end{table}

We first split the 40,000 $({\bm \theta}_i,{\bf x}_i)$ samples in training and validation sets, with a 90/10 split ratio. Then, we train the MAF architecture $q_\phi({\bm \theta}|{\bf x})$ with parameters $\phi$ to approximate the posterior distribution $p({\bm \theta}|{\bf x})$. The training is performed by minimizing the Kullback-Leibler divergence between the two distributions, which is equivalent to maximizing the score function
\begin{equation}
    S(\phi) = \sum_{i\in T}\log q_\phi({\bm \theta}_i|{\bf x}_i)
\end{equation}
where the sum is performed over the training set $T$. We use the {\sc Adam} \citep{Kingma2014} algorithm to maximize the score function. At the end of each training epoch, we evaluate the score function on the validation set. To avoid overfitting, we stop the training procedure when the validation score function fails to increase after 20 epochs. 

Both the MAF architecture and the training procedure have hyperparameters that can be chosen to maximize the score function. They are the number of affine transforms in the MAF, the number of blocks and hidden units of the multi-layer perceptrons (MLP), the dropout probability, the learning rate, and the batch size. We sample the hyperparameter space (with range for each parameter listed in Table~\ref{tab:hyperparams}) using the {\sc Optuna} \citep{Akiba2019} software. For each $M_\ell(k)$ considered, we train at least 300 different architectures. In our setup, {\sc Optuna} chooses the hyperparameters of the first 100 trials randomly, then it uses the Tree-structured Parzen Estimator (TPE) to sample the hyperparameter space and determine the value of the hyperparameters of the following architecture. Since ensembling NDEs with different initializations and architectures generally improves robustness \citep{Lakshminarayanan2016,Alsing2019,Yao2023}, we select the 5 best ones and take their ensemble to obtain a final, more accurate, model $q_\phi^{\rm ens}({\bm \theta}|{\bf x})$. To select the 5 best architectures, we first pick the 10 with the largest validation score, then, among those, we select 5 with validation rank distributions for the cosmological parameters $\Omega_{\rm m}$ and $\sigma_8$ that are the closest to a uniform distribution (see Section~\ref{sec:validation} for more details on rank distributions). We select the best NDEs based on rank statistics, in addition to the score function, to allow us to discard an architecture if it gives overconfident or biased posteriors for $\Omega_{\rm m}$ and $\sigma_8$. The presence of NDEs with high validation scores and non-uniform rank distribution may indicate the need to further explore the hyperparameter space and train more than 300 architectures. Given the large number of different summary statistics we want to explore (eight individual marked power spectra and some combination of joint marked power spectra), we limit the number of architectures that we train for each statistic due to computing resources. 

We apply this procedure for all the $M_\ell(k)$ listed in Table~\ref{tab:Mk_models}. 
For simplicity and clarity, in the remaining part of the paper we will focus only on the $q_\phi^{\rm ens}({\bm \theta}|{\bf x})$ obtained using as summary statistics three specific marked power spectra that give robust and interesting constraints on cosmology. They are $M^B_\ell$, which gives the best constraint on $\sigma_8$, $M^G_\ell$, which provides the best constraint on $\Omega_{\rm m}$, and $M^C_\ell$. We also consider the concatenation of pairs of these marked power spectra as summary statistics, and we train, validate, and deploy $q_\phi^{\rm ens}({\bm \theta}|{\bf x})$ in these cases.

\subsection{Posterior Validation}

\label{sec:validation}

\begin{figure}
\begin{center}
    \includegraphics[width=1\columnwidth]{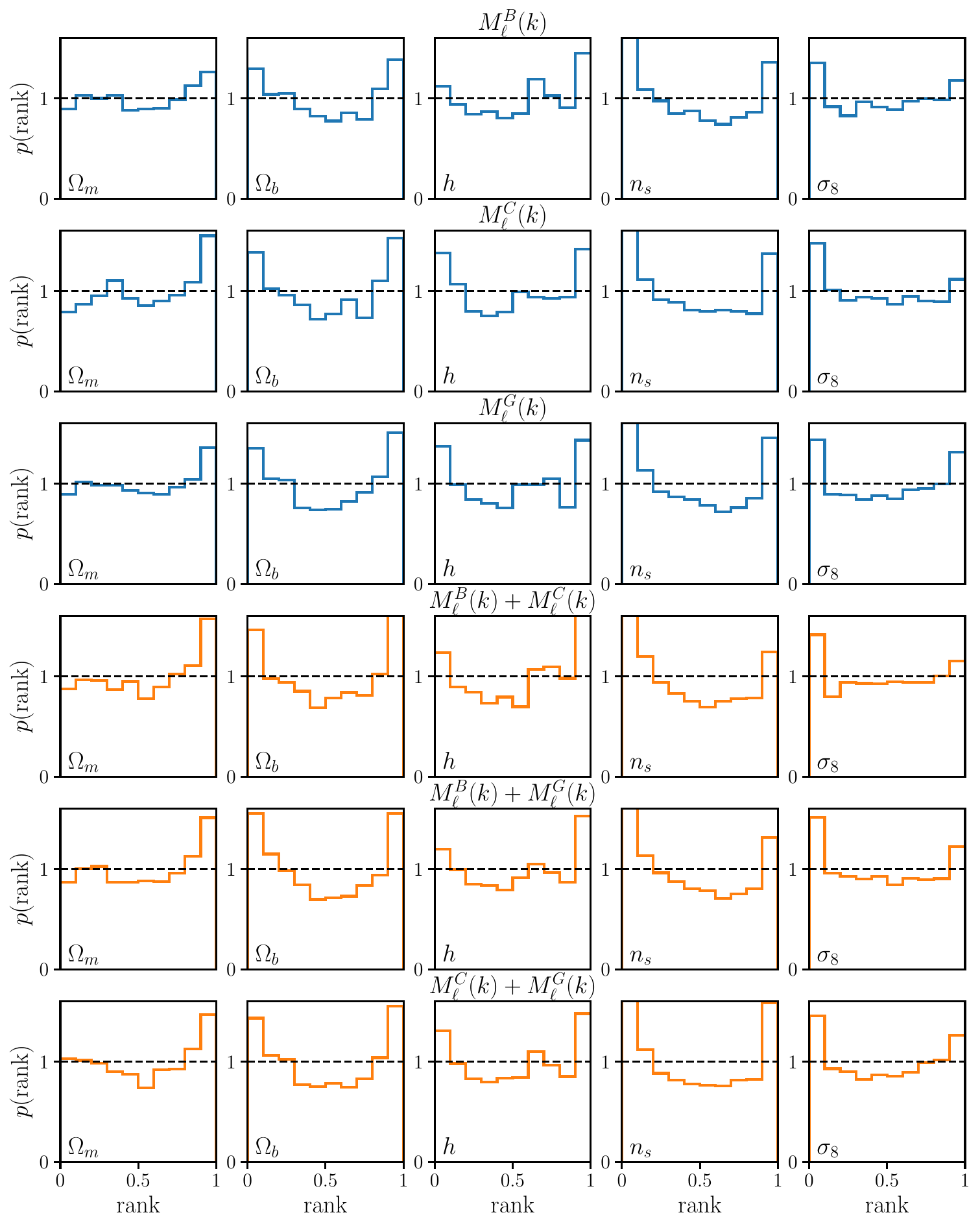}
    \caption{\label{fig:Mk_rank} Rank distribution derived from the validation set to assess the accuracy of the estimated posteriors. Each column shows the results for a particular cosmological parameter and each row for a particular summary statistic. Results for a single $M_\ell(k)$ are displayed in blue and for the concatenation of two marked power spectra in orange. The rank statistics should be uniformly distributed for an ideal posterior estimate (black dashed).
    Our results indicate that the posteriors tend to be overconfident, especially for $\Omega_{\rm b}$, $h$, and $n_s$. 
}
\end{center}
\end{figure}
Before applying our $M_\ell(k)$ posterior estimator $q_\phi^{\rm ens}({\bm \theta}|{\bf x})$ to our observations, we test its accuracy and robustness.  

First, we perform simulation-based calibration \citep[SBC;][]{Talts2020} for the $\Lambda$CDM parameters to check if $q_\phi^{\rm ens}({\bm \theta}|{\bf x})$ accurately estimates the true posterior. We employ the validation set used for the early-stop procedure during training. We apply $q_\phi^{\rm ens}({\bm \theta}|{\bf x})$ to the summary statistic ${\bf x}_i=\left\{M_\ell(k),\bar{n}_g\right\}_i$ of each validation sample $i$ to infer their posterior $q_\phi^{\rm ens}({\bm \theta}|{\bf x}_i)$. Then, from each $q_\phi^{\rm ens}({\bm \theta}|{\bf x_i})$, we draw $10,000$ independent samples $\hat{{\bm \theta}}_{i}$, and use them to compute the rank of each component of ground truth ${\bm \theta}_i$ within the 1D posterior distribution. Once we have the rank for each validation data point, we compute the normalized distribution of the ranks. Figure~\ref{fig:Mk_rank} shows the rank distribution for the 5 cosmological parameters for different marks models.  
The rank distributions for $M^B_\ell(k<0.5~h{\rm Mpc}^{-1})$, $M^C_\ell(k<0.5~h{\rm Mpc}^{-1})$, and $M^G_\ell(k<0.5~h{\rm Mpc}^{-1})$ are displayed in blue on top, and for their combinations in orange on bottom. A uniform (flat) distribution indicates that the estimated posterior distribution $q_\phi^{\rm ens}({\bm \theta}|{\bf x})$ is compatible with the true posterior. In our case, the rank distributions of $\Omega_{\rm b}$, $h$, and $n_s$ present a $\cup$-like shape, indicating that the marginalized posteriors for these parameters are overconfident. These results are not ideal, but they are not too worrisome since the posterior distribution of these parameters is dominated by the prior range (see Section~\ref{sec:results}). The rank distributions of $\sigma_8$ are almost uniform, indicating accurate marginalized posteriors for this parameter. Those of $\Omega_{\rm m}$ are also almost uniformly distributed but in some cases, especially for model C, the value of the distribution somewhat increases with the rank, showing a slight indication of bias.  

\begin{figure*}
\begin{center}
    \includegraphics[width=1.7\columnwidth]{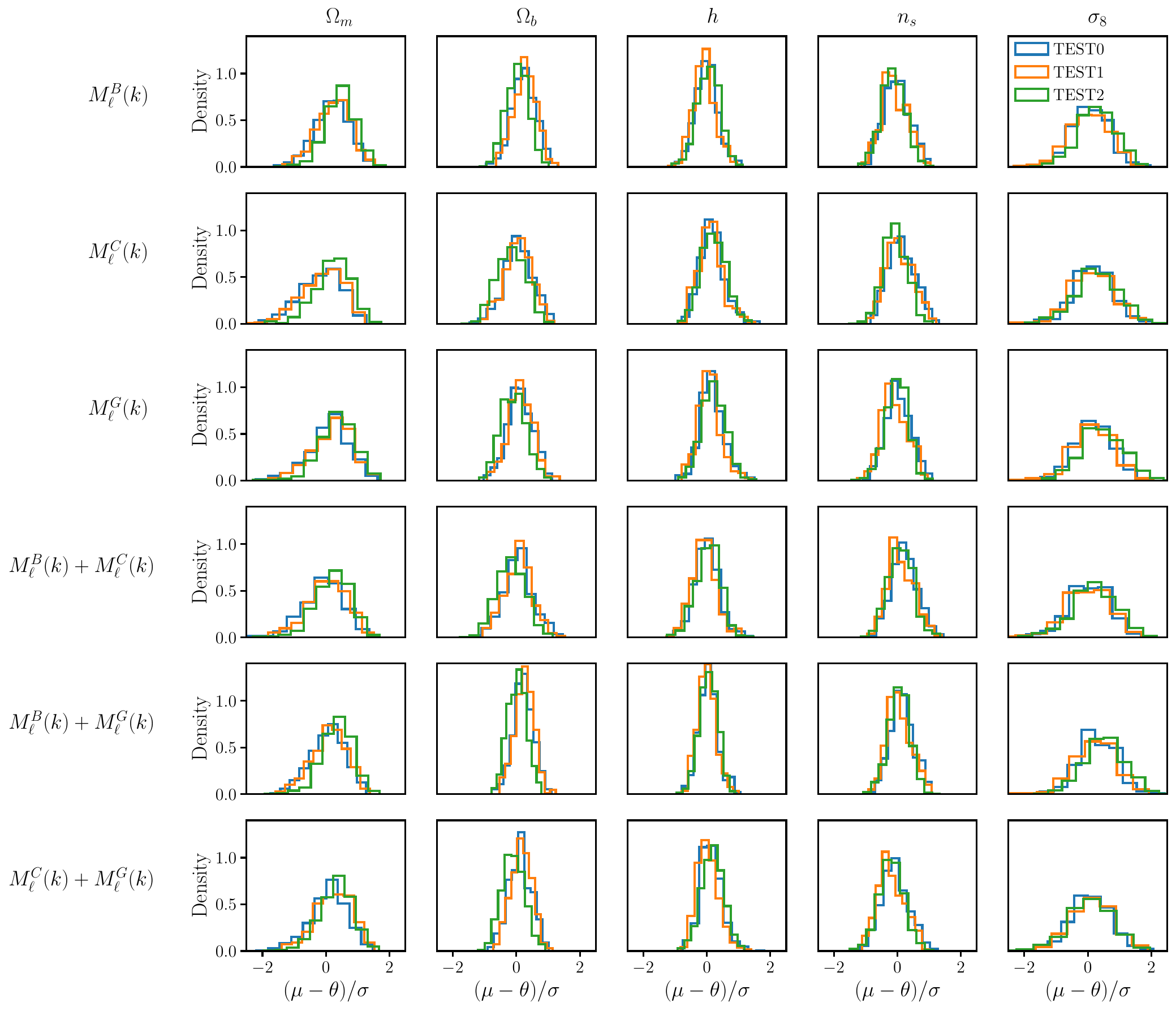}
    \caption{\label{fig:Mk_cstat} Distributions of the differences between the posterior mean $\mu$ and the true parameter $\theta_{\rm fid}$ normalized by the posterior standard deviation $\sigma$, for each cosmological parameter (columns). Each row shows the results obtained with different summary statistics (a single or concatenation of two marked power spectra) for the 
    {$\mathtt{TEST0}$} (blue), {$\mathtt{TEST1}$} (orange), and {$\mathtt{TEST2}$} (green) mock challenge catalogs. 
    In each panel, the distributions of the 3 test sets overlap, indicating that our cosmological inference is robust.  
}
\end{center}
\end{figure*}

Second, we test the robustness (against changes in the forward model) of $q_\phi^{\rm ens}({\bm \theta}|{\bf x})$ by applying it to the \simbig~mock challenge data described in Section~\ref{sec:mock_challenge_data}. For each catalog in {$\mathtt{TEST0}$}, {$\mathtt{TEST1}$}, and {$\mathtt{TEST2}$} we sample the posterior distribution inferred by our pipeline and compute the mean $\mu$ and standard deviation $\sigma$ of the marginalized posterior distribution for each cosmological parameter. Then, we calculate the distribution of the quantity $(\mu-\theta)/\sigma$ for each test set. Figure~\ref{fig:Mk_cstat} shows the results obtained with different summary statistics (rows) and for different cosmological parameters (columns) and different test sets (color-coded). We can interpret $\mu$ as a compressed summary statistic and its distribution as the likelihood $p(\mu|\theta)$. When the cosmological inference is robust to the modeling choices in our forward model ($N$-body solver, halo definition, halo-galaxy connection), we expect the distribution of $(\mu-\theta)/\sigma$ to be similar among the different test sets\footnote{We consider the quantity $\mu-\theta$ instead of $\mu$ because the simulations in {$\mathtt{TEST0}$} and {$\mathtt{TEST1}$} are at the same cosmology while those in {$\mathtt{TEST2}$}
are at a different one.}. We see that the distributions for {$\mathtt{TEST0}$}, {$\mathtt{TEST1}$}, and {$\mathtt{TEST2}$} are in good agreement with each other in all the panels. 
This indicates that $q_\phi^{\rm ens}({\bm \theta}|{\bf x})$ for each of the six different summary statistics ${\bf x}$ are sufficiently robust against changes in the choices of the forward model.

\section{Results}
\label{sec:results}

\begin{figure*}
\begin{center}
    \includegraphics[width=0.9\textwidth]{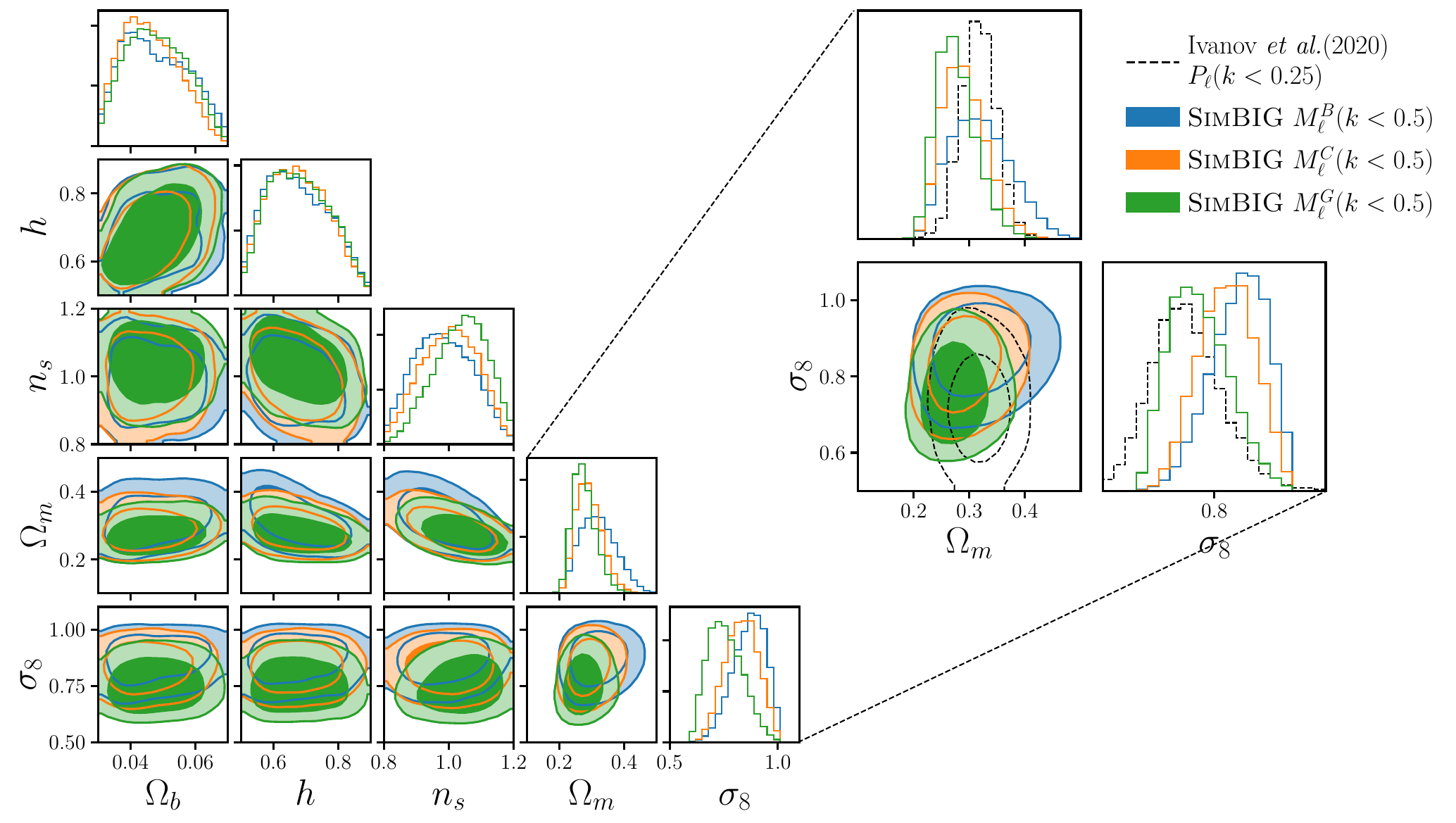}
    \caption{\label{fig:Mk_posteriors}
{\em Left}: Posterior distributions for all the $\Lambda$CDM cosmological parameters inferred from the BOSS CMASS SGC sample using the marked power spectra $M^B_\ell(k<0.5~h {\rm Mpc}^{-1})$ (blue), $M^C_\ell(k<0.5~h {\rm Mpc}^{-1})$ (orange), and $M^G_\ell(k<0.5~h {\rm Mpc}^{-1})$ (green) with \simbig. The contours represent the 68 and 95 percentiles. {\em Right}: Posterior distributions for $\Omega_{\rm m}$ and $\sigma_8$. For comparison, we display the posteriors from the PT-based $P_\ell(k<0.25~h {\rm Mpc}^{-1})$ analysis (black dashed) in \cite{Ivanov2020}.}
\end{center}
\end{figure*}

After testing the accuracy and robustness of our inference pipeline, we apply $q_\phi^{\rm ens}({\bm \theta}|{\bf x})$ to the BOSS CMASS SGC observational data. In Section \ref{sec:result_1M} we present the posterior distributions for the cosmological parameters obtained using individual marked power spectra as summary statistics, while in Section \ref{sec:result_2M} we show the results using concatenations of two marked power spectra.

\subsection{Individual marked power spectrum}
\label{sec:result_1M}

\begin{table*}
    \def\arraystretch{1.3}
   \centering
   \begin{tabular}{|c|ccccc|}
       \hline
       Statistics & $\Omega_{\rm b}$ & $h$ & $n_s$ & $\Omega_{\rm m}$ & $\sigma_8$ \\
       \hline
       $M^B_\ell$ & $0.047^{+0.012}_{-0.009}$ & $0.668^{+0.113}_{-0.092}$& $0.981^{+0.097}_{-0.090}$& $0.317^{+0.059}_{-0.052}$& $0.873^{+0.069}_{-0.086}$ \\
       $M^C_\ell$ &$0.045^{+0.010}_{-0.008}$& $0.672^{+0.101}_{-0.090}$& $1.005^{+0.087}_{-0.094}$& $0.289^{+0.044}_{-0.036}$& $0.835^{+0.076}_{-0.080}$ \\
       $M^G_\ell$ &$0.048^{+0.010}_{-0.009}$& $0.676^{+0.106}_{-0.092}$& $1.042^{+0.077}_{-0.092}$& $0.273^{+0.040}_{-0.030}$& $0.753^{+0.091}_{-0.074}$ \\
       \hline
       $M^B_\ell+M^C_\ell$ &$0.042^{+0.010}_{-0.007}$& $0.654^{+0.103}_{-0.081}$& $1.025^{+0.085}_{-0.098}$& $0.287^{+0.051}_{-0.041}$& $0.820^{+0.078}_{-0.083}$ \\
       $M^B_\ell+M^G_\ell$ &$0.048^{+0.012}_{-0.010}$& $0.683^{+0.110}_{-0.102}$& $1.022^{+0.086}_{-0.097}$& $0.286^{+0.054}_{-0.039}$& $0.839^{+0.076}_{-0.086}$ \\
       $M^C_\ell+M^G_\ell$ &$0.047^{+0.011}_{-0.009}$& $0.674^{+0.109}_{-0.096}$& $0.998^{+0.084}_{-0.086}$& $0.291^{+0.048}_{-0.039}$& $0.777^{+0.077}_{-0.071}$ \\
       \hline
   \end{tabular}
   \caption{List of the 50, 16, and 84th percentile constraints ($68\%$ CL) on the $\Lambda$CDM cosmological parameters inferred from the BOSS CMASS SGC sample using the marked power spectra $M^B_\ell(k<0.5~h {\rm Mpc}^{-1})$, $M^C_\ell(k<0.5~h {\rm Mpc}^{-1})$, $M^G_\ell(k<0.5~h {\rm Mpc}^{-1})$, and their combinations with the \simbig~framework.}
   \label{tab:constraints}
\end{table*}

\begin{figure*}
\begin{center}
\subfloat[mark configurations B and C]{%
  \includegraphics[width=0.3\textwidth]{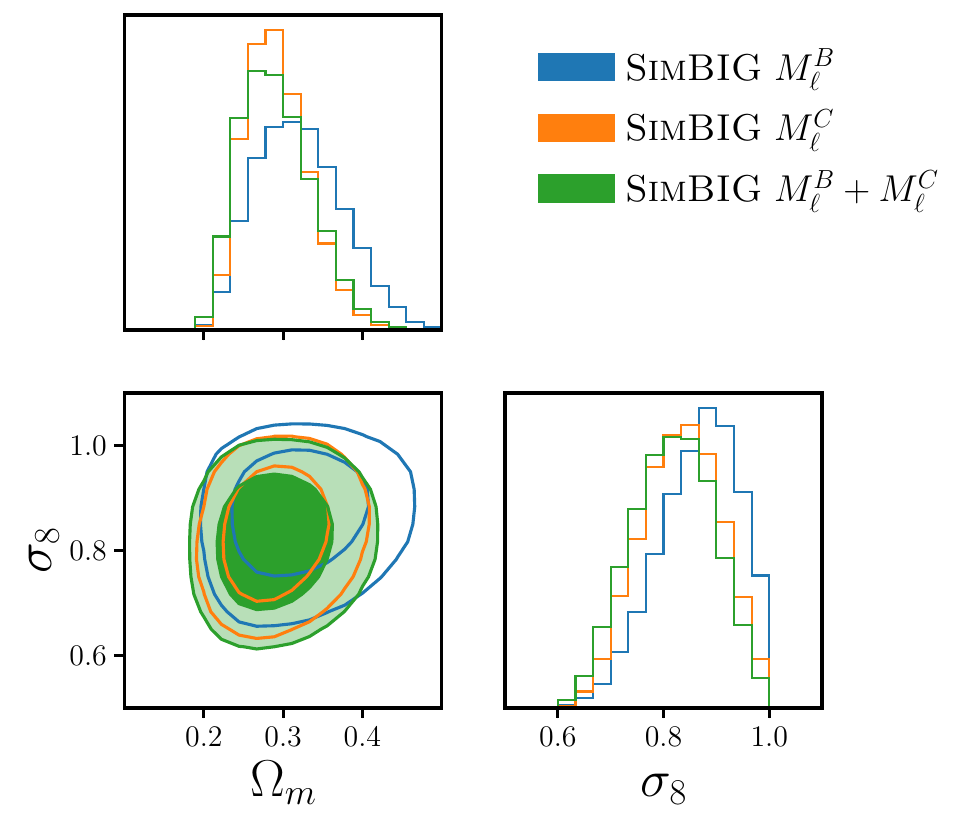}%
  
}
\subfloat[Mark configurations B and G]{
  \includegraphics[width=0.3\textwidth]{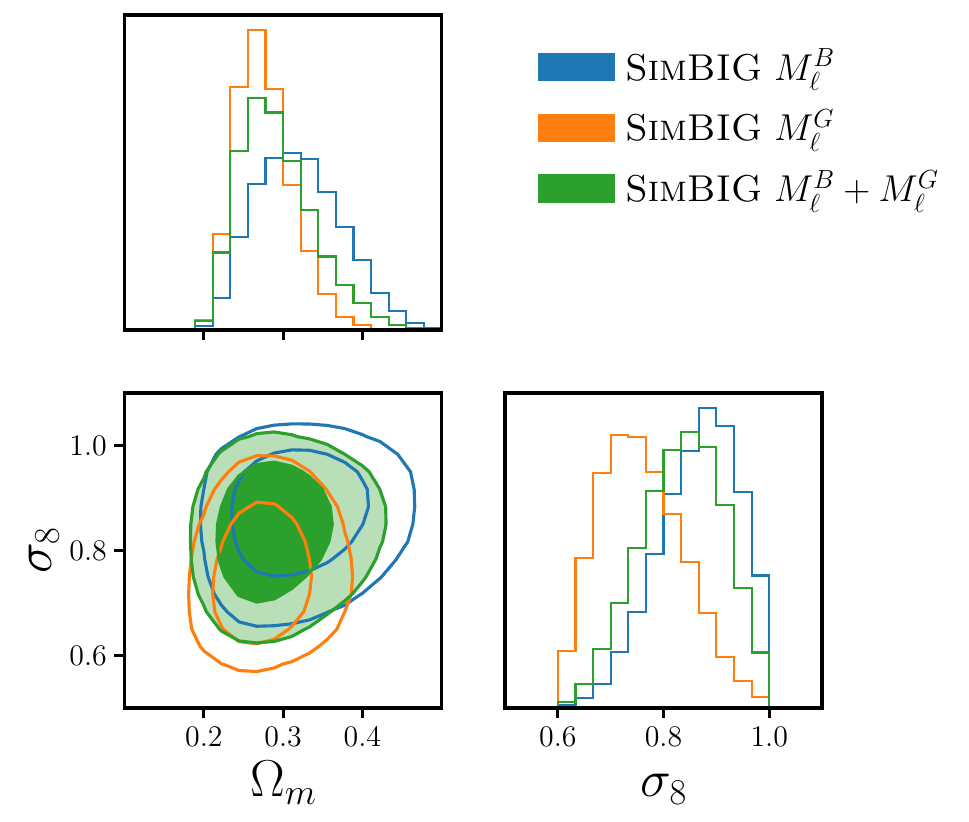}
}
\subfloat[Mark configurations C and G]{%
  \includegraphics[width=0.3\textwidth]{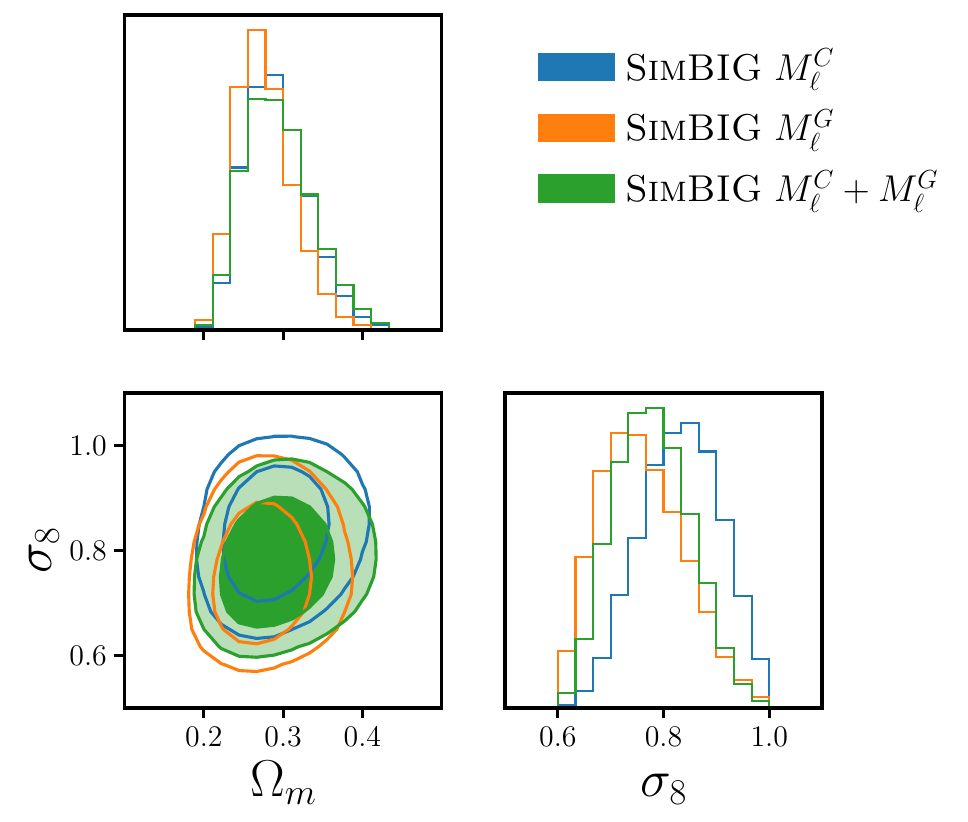}
}
\caption{\label{fig:Mk_posteriors_2M} Posterior distributions for $\Omega_{\rm m}$ and $\sigma_8$ obtained from the BOSS CMASS SGC sample using a single marked power spectrum (blue and orange) and their combination (green) with \simbig. From left to right, we use (a) $M^B_\ell(k<0.5~h {\rm Mpc}^{-1})$ and $M^C_\ell(k<0.5~h {\rm Mpc}^{-1})$, (b) $M^B_\ell(k<0.5~h {\rm Mpc}^{-1})$ and $M^G_\ell(k<0.5~h {\rm Mpc}^{-1})$, and (c) $M^C_\ell(k<0.5~h {\rm Mpc}^{-1})$ and $M^G_\ell(k<0.5~h {\rm Mpc}^{-1})$. The posteriors obtained with two marked power spectra are consistent with those from the single $M_\ell(k)$, but not necessarily tighter. Only the case $M^C_\ell+ M^G_\ell(k)$ gives a tighter constrain on $\sigma_8$ than the instances where the two marked power spectra are used singularly. 
}
\end{center}
\end{figure*}

Figure~\ref{fig:Mk_posteriors} shows the cosmological posterior distributions obtained using the three selected marked power spectra, $M^B_\ell(k)$ in blue, $M^C_\ell(k)$ in orange, and $M^G_\ell(k)$ in green. The contours represent the 68 and 95 percentiles and the ranges of the panels match the prior, except for $\sigma_8$ where the range has been extended for better visualization. The 50, 16, and 84th percentile constraints on the cosmological parameters are listed in the top part of Table~\ref{tab:constraints}. 

Our analysis indicates that $M^B_\ell(k)$, $M^C_\ell(k)$, and $M^G_\ell(k)$ set similar constraints on the parameters $\Omega_{\rm b}$ and $h$, which are prior dominated. These parameters are not well constrained by galaxy clustering probes alone. Instead, they are best constrained by Cosmic Microwave Background data via the combination $\omega_{\rm b}=\Omega_{\rm b}h^2$, and the Big Bang Nucleosynthesis (BBN) predictions. Most galaxy clustering analyses apply a BBN prior in their inference pipeline \citep[e.g.,][]{Ivanov2020}. 
Our results show that the marginalized posteriors of $n_s$ are also broad and dominated by the priors. 

The cosmological parameters related to the growth of structure, $\Omega_{\rm m}$ and $\sigma_8$, are those better constrained by galaxy clustering analyses. Using $M_\ell(k)$  applied to the \simbig~pipeline, we obtain the following results: $\Omega_{\rm m}=0.317^{+0.059}_{-0.052}$ and $\sigma_8=0.873^{+0.069}_{-0.086}$ for $M^B_\ell(k)$, $\Omega_{\rm m}=0.289^{+0.044}_{-0.036}$ and $\sigma_8=0.835^{+0.076}_{-0.080}$ for $M^C_\ell(k)$, and $\Omega_{\rm m}=0.273^{+0.039}_{-0.031}$ and $\sigma_8=0.750^{+0.091}_{-0.074}$ for $M^G_\ell(k)$. The 1D-marginalized posteriors for $\Omega_{\rm m}$ obtained with the three different marked models are consistent, and that derived using the mark model $G$ is the tightest. Model $G$ has the smallest value for the scale $\rm{R}$ among the three models, indicating that $\delta_{\rm R}$ and the mark are defined only by the closest galaxies. Moreover, it has exponent $p$ equal to 0.5, meaning that low-density regions are up-weighted compared to high-density ones, but with less importance than, for example, in model $B$, where $p=1$. The marginalized posteriors for $\sigma_8$ exhibit a similar width and they are slightly off-centered, with the median of distribution from model $B$ being 1-sigma away from the median of the posterior from model $G$, giving consistent results. 

We compare our \simbig~$M_\ell(k)$ posteriors to the PT analysis with the power spectrum multiples $P_\ell(k<0.25~h{\rm Mpc}^{-1})$ from \cite{Ivanov2020} in the right inset in Figure~\ref{fig:Mk_posteriors}. The inset zooms in the $\Omega_{\rm m}-\sigma_8$ plane and additionally displays the posteriors from the PT $P_\ell(k<0.25~h{\rm Mpc}^{-1})$ analysis as a black dashed line. The \simbig~$M^B_\ell(k)$, $M^C_\ell(k)$ and $M^G_\ell(k)$ constraints on $\sigma_8$ are $1.20\times$, $1.19\times$, and $1.12\times$ tighter than those from the PT analysis, respectively. This gain is likely due to the inclusion of nonlinear scales and non-Gaussian information in the analysis. However, compared to the \simbig~$P_\ell(k)$ analysis performed in \citepalias{simbig_letter}, where all scales up to $k<0.5~h{\rm Mpc}^{-1}$ were included, we do not find any improvement in constraining $\sigma_8$ when using these mark configurations, but a $1.2\times$ and $1.3\times$ tighter constraint on $\Omega_{\rm m}$ when using $M^C_\ell$ and $M^G_\ell$.

\subsection{Two joint marked power spectra}
\label{sec:result_2M}

Fisher analyses that used marked power spectra of the matter field \citep{Massara2021} and of the galaxy field \citep{Massara2023} showed that using two or more marked power spectra jointly could improve the cosmological constraints obtained with the same marked power spectra but used individually. We apply the $q_\phi^{\rm ens}$ trained using the concatenation of two marked power spectra to CMASS SGC, and obtain the results listed in the lower part of Table~\ref{tab:constraints}. The constraints on $\Omega_{\rm b}$, $h$, and $n_s$ do not improve significantly when considering two joint marked power spectra instead of one, except for the combination $B+C$. Even in that case, a fair comparison is difficult to make, since all the $q_\phi^{\rm ens}$ are overconfident to different extents when predicting the posteriors of these cosmological parameters (see Figure~\ref{fig:Mk_rank}). 

Figure~\ref{fig:Mk_posteriors_2M} shows the posterior distributions obtained with two marked power spectra (green) and a single marked power spectrum (orange and blue) in the $\Omega_{\rm m}-\sigma_8$ plane. In general, the 1D posterior from the concatenation of two $M_\ell(k)$ appears to be broader than those from the same marked power spectra used individually while being placed halfway in between their posteriors, which indicates a consistency throughout the inference pipelines. There is an indication of preference for the use of two marked power spectra instead of one only in the case $C+G$ and to constrain $\sigma_8$. In that case, the constraint coming from the two joint marked power spectra is about $1.1\times$ tighter than those from the individual $M_\ell(k)$.

Overall, there is no improvement in cosmological constraining power when using two marked power spectra jointly instead of one. This is in disagreement with predictions from Fisher forecast and with the intuition that different mark configurations may contain complementary information and exhibit different parameter degeneracies. Similar marked power spectra analyses applied to galaxy samples with different characteristics, such as number density, bias, etc, might be needed to understand the nature of our findings.

\section{Discussion}
\label{sec:discussion}

\begin{figure}
\begin{center}
    \includegraphics[width=0.8\columnwidth]{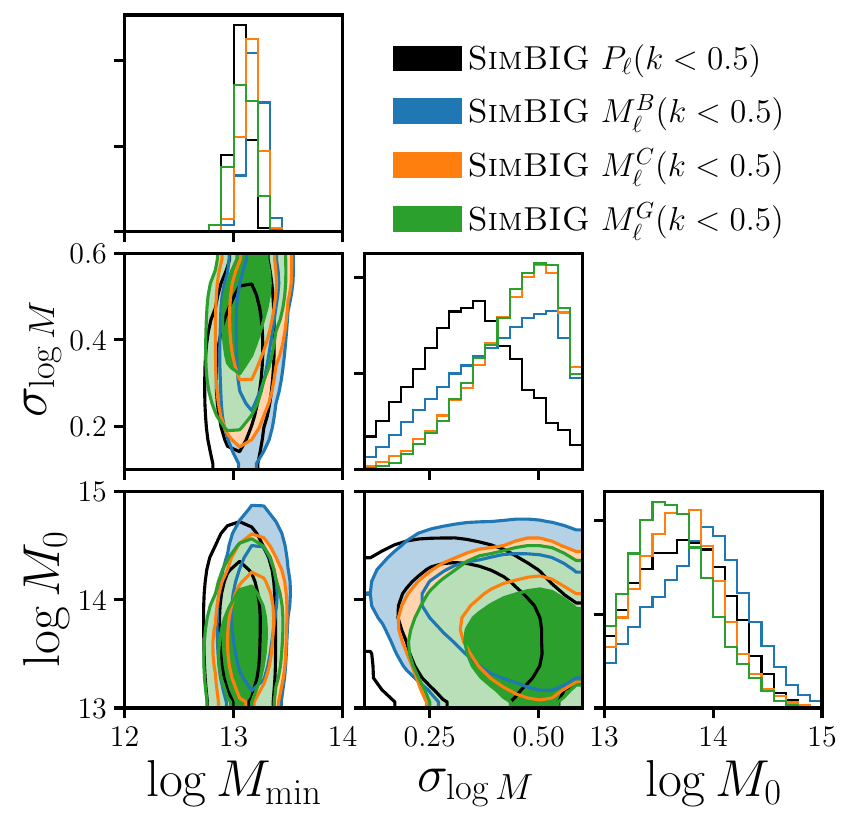}
    \caption{\label{fig:Mk_3HOD_post}
 Posterior distributions for the HOD parameters $\log M_{\rm min}$, $\sigma_{\log M}$, and $\log M_0$ inferred from the BOSS CMASS SGC sample using the marked power spectra $M^B_\ell(k<0.5~h {\rm Mpc}^{-1})$ (blue), $M^C_\ell(k<0.5~h {\rm Mpc}^{-1})$ (orange), $M^G_\ell(k<0.5~h {\rm Mpc}^{-1})$ (green), and the standard power spectrum $P_\ell(k<0.5~h {\rm Mpc}^{-1})$ with \simbig. The contours represent the 68 and 95 percentiles. }
\end{center}
\end{figure}

The Fisher analysis performed in \cite{Massara2023} showed that some marked power spectra are better at constraining $\sigma_8$ than the standard power spectrum. This was already anticipated in \cite{White2009}, where the authors showed that marked correlation functions are expected to break the degeneracy between the HOD parameters and $\sigma_8$ present in the correlation function, or $P_\ell(k)$. This idea was corroborated by the Fisher analysis in \cite{Massara2023}, which showed that $M_\ell(k)$ improves the $P_\ell(k)$ HOD constraints, and in particular those that describe the central galaxies, $\log M_{\rm min}$ and $\sigma_{\log M}$, and $\log M_0$, which describes the minimum halo mass hosting a satellite galaxy. The posterior distributions for those three HOD parameters inferred from the observed CMASS SGC sample are shown in Figure~\ref{fig:Mk_3HOD_post}, where results obtained using the three selected marked power spectra are indicated in blue, orange, and green, and those from the \simbig~$P_\ell(k)$ are shown in black for comparison. Our \simbig~analysis hints to an improvement in constraining $\sigma_{\log M}$ and $\log M_0$ when using these marked power spectra compared to $P_\ell(k)$, with the 1D-posteriors for $\sigma_{\log M}$ being skewed towards large values. Nevertheless, the improvement in constraining $\sigma_8$ is not as large as expected in the Fisher analysis, particularly when compared to the \simbig~$P_\ell(k)$ results. There are at least two possible motivations for this finding. 

First, marked power spectra applied to galaxy survey data could carry less cosmological information than what was forecast in Fisher analyses with $N$-body simulations. In general, this could be due to the presence of survey geometry and observational systematics in the observed dataset, and the assumption of a Gaussian form for the posterior distribution in the Fisher analysis. Moreover, in this specific comparison, the Fisher analysis in \cite{Massara2023} has been developed using the standard 5-parameter HOD framework, while the \simbig~forward model employs a more sophisticated HOD model with 9 parameters, and lower resolution $N$-body simulations of the Quijote suite, which required a large number of satellite galaxies to reproduce an SDSS-like number density. The large satellite fraction impacted the multipoles of the standard power spectrum in the following way: the quadrupole dropped to negative values already on large scales and the hexadecapole was positive on small scales (see figure 23 in~\cite{Hou2023Fisher}), while it is close to zero in the CMASS sample. Similarly, the peculiar HOD settings can have impacted marked power spectra. 

Second, the CMASS SGC sample may not be the ideal galaxy sample to exploit marked power spectra. The CMASS SGC sample is not very dense, $\bar{n}\sim 2 \cdot 10^{-4} ~h^{-3}{\rm Mpc}^3$, and does not include galaxies in small halos---CMASS galaxies reside in halos with $M_h > 10^{13}h^{-1}M_\odot$. On average, halos in low-density regions are smaller than those in high-density environments, thus the CMASS sample lacks galaxies in voids. Marked power spectra that up-weight low-density regions are built to boost the signal from those galaxies that are not observed in CMASS. Additional marked power spectra analyses on denser samples will need to be carried out to assess if any or both of these reasons are responsible for our findings. 

Due to computing costs, our analysis could not be performed using a very large number of different mark configurations. We chose to investigate those that were more suitable for the characteristics of the CMASS sample, where our choice of what is suitable was informed by the following considerations. Positive values for the exponent $p$ correspond to $M_\ell(k)$ that up-weight galaxies in low-density regions, and, the larger the value of $p$, the stronger this up-weight is. We might think that a large $p$ is desirable since it boosts even more the information from voids, but it is not always the case.  
If the sample includes a low number of galaxies in voids, their low number increases the {\it mark-dependent} Poisson shot noise in $M_0(k)$. 
More specifically, while the Poisson shot noise of the power spectrum is $1/\bar{n}$, where $\bar{n}$ is the average galaxy density, the Poisson shot noise of the marked power spectrum is $\overline{m^2}/(\bar{m}^2\bar{n})$, where $\bar{m}^2$ and $\overline{m^2}$ are the average of the mark field and the squared mark field, with the average weighted by the galaxy number density field (see \cite{Philcox2020} for the derivation). Therefore, the Poisson shot noise of $M_0(k)$ depends on the type of galaxies in the sample and their environmental density via the mark. Configurations $A$ and $D$ (in blue and red in Figure~\ref{fig:Mk_measured}) have exponent $p=2$, the largest considered in this work. Because of the characteristics of the CMASS sample, these configurations exhibit a large shot noise, which shifts the monopoles of $M^A$ and $M^D$ towards higher values. This can be understood by the high values of their monopoles on small scales compared to other mark configurations. Likely because of large shot noise, the marked power spectra $M^A$ and $M^D$ gave poorer constraints on the cosmological parameters compared to other mark configurations with smaller values for $p$, and the degradation was evident when looking at the growth of structure parameters $\Omega_{\rm m}$ and $\sigma_8$. When a denser sample with lower galaxy bias including more faint galaxies that are in voids will be available, using models with $p=2$ might be beneficial and allow us to obtain better results than employing models with $p<2$. 

The choice of the scale $\rm{R}$ is also important. A very large $\rm{R}$ gives a smoothed overdensity galaxy field $\delta_{\rm R}$ that is close to zero everywhere, a corresponding mark equal to unity, and a marked power spectrum that reduces to the standard power spectrum. On the other hand, a very small $\rm{R}$ results in an inaccurate prediction for $\delta_{\rm R}$ because of the discrete nature of galaxies. The sparser the galaxy sample, the larger the minimum scale $\rm{R}$ that gives a reasonable estimation of the smoothed galaxy field. In our analysis, we considered values of $\rm{R}$ down to $10~h^{-1}M{\rm pc}$, which corresponds to a sphere where on average there are roughly 0.6 galaxies of the CMASS sample. The mark model considering $\rm{R}=10~h^{-1}{\rm Mpc}$ is $H$, which has $p=1$ and $\delta_s=0.1$. This case passed all the validation tests described in Section~\ref{sec:validation} and gave the following constraints when applied to CMASS SGC: $\Omega_{\rm m}=0.227^{+0.047}_{-0.022}$, $\sigma_8=0.727^{+0.104}_{-0.076}$, $n_s=1.097^{+0.062}_{-0.091}$. The constraint on $\Omega_{\rm m}$ is the most stringent among all the ones obtained with a single $M_\ell(k)$, however, the 1D-posterior distribution of $\Omega_{\rm m}$ hits the lower-bound of the prior, making the analysis less reliable than those performed with other mark configurations. Similarly, the 1D-posterior of $n_s$ is near the upper bound of the $n_s$ prior. This hints at the possibility that $\rm{R}=10~h^{-1}$Mpc is too small for the considered galaxy sample. 

Many ongoing and upcoming galaxy surveys will observe samples that are denser than the BOSS survey. The DESI Bright Galaxy Survey (BGS) will produce a very detailed map of the large-scale structure containing about $10$ million galaxies and spanning $14,000$ deg$^2$ at low redshift, $z<0.6$ \citep{Hahn2023}, while Roman will observe a galaxy sample with similar density and volume but at much high redshift, which will include 10 million H$\alpha$ emitters within $2,000$ deg$^2$ at redshift $1<z<2$ \citep{Wang2022}. Therefore, these surveys will produce samples that are about an order of magnitude denser than our BOSS CMASS SGC subsample while having a volume comparable to the full BOSS sample. Another interesting galaxy catalog with larger volume but lower density will be produced by Euclid, which will observe about 30 million H$\alpha$ emitters within $15,000$ deg$^2$ in the redshift range $0.9<z<1.8$ \citep{Laureijs2011}. That sample will be roughly 4 times denser than our BOSS CMASS subsample and span a volume 6 times larger. These reach data sets will allow us to probe different epochs and regimes of the large-scale structure, and they will likely be ideal for performing SBI cosmological analyses with marked power spectra.

\section{Conclusions}
\label{sec:conclusions}
We present the \simbig~cosmological constraints obtained by analyzing different marked power spectra measured from a subset of the CMASS SGC galaxy sample. The \simbig~framework uses state-of-art $N$-body simulations and HOD scheme and includes survey geometry, masks, and fiber collision effect to forward modeling the galaxy field. It provides the training, validation, and test sets to perform SBI and learn NDEs with normalizing flows, allowing us to perform the cosmological analysis up to small nonlinear scales and without assuming a Gaussian form for the likelihood, which is directly learned from the forward model.  

Previous works have used \simbig~to perform the cosmological analysis with the CMASS SGC sample using power spectrum, bispectrum, skew spectra, wavelet scattering transform, and CNNs. We introduce an additional statistic, the marked power spectrum, in the pipeline. Multiple marked power spectra can be measured from the same galaxy sample varying the mark parameters $\rm{R}$, $p$, and $\delta_s$. We consider seven different mark configurations and, for each of them, obtain an NDE that describes their posterior distribution and passes a validation test inspecting their accuracy for at least $\Omega_{\rm m}$ and $\sigma_8$, which are the parameters that are not prior dominated. 

\simbig~provides additional galaxy catalogs built with different forward models to perform a mock challenge aimed at testing the robustness of the inferred posterior distributions. After checking that our NDEs pass these tests, we apply them to the marked power spectra measured from the CMASS SGC sample. Among all the mark configurations, we select those three that give the best and most reliable constraints on the growth of structure parameters. We use pairs of these marked power spectra as summary statistics and train, validate, and test new NDEs to describe their posterior distributions. 

Overall, considering an individual or the concatenation of two marked power spectra, the best constraints on the growth of structure parameters are $\Omega_m=0.273^{+0.040}_{-0.030}$ and $\sigma_8=0.777^{+0.077}_{-0.071}$. The $\sigma_8$ constraint is $1.2\times$ tighter than that derived from a PT power spectrum analysis using the same dataset, while we do not see any improvement in constraining $\Omega_m$. 

The fact that (1) the \simbig~$M_\ell(k)$ analyses exhibit improved constraints only on $\sigma_8$ when compared to a PT $P_\ell(k)$ analysis and (2) the improvement is similar to that achieved with \simbig~$P_\ell(k)$ suggest that marked power spectra that up-weight low-density regions are not effective as expected at retrieving cosmological information from the CMASS SGC sample. We suspect this is due to the low galaxy number density and the high galaxy bias characterizing this population. These characteristics indicate that there is no information coming from low-density regions in this sample. Further investigation involving SBI on denser and less biased galaxy samples is required to determine if this reasoning is valid. 

In the future, it will be interesting to perform SBI with marked power spectra from denser samples to constrain not only the $\Lambda$CDM parameters but also physics beyond the $\Lambda$CDM model. The effects of massive neutrinos and modified gravity are pronounced in low-density regions. Therefore, marked power spectra that up-weight low-density regions are expected to be particularly powerful in probing the neutrino mass scale and constraining gravity.

\section*{Acknowledgements}
It's a pleasure to thank Mikhail M. Ivanov for providing us with the posteriors used for comparison.
CH was supported by the AI Accelerator program of the Schmidt Futures Foundation.

\bibliography{Mk_bib}{}

\begin{thebibliography}{}
\expandafter\ifx\csname natexlab\endcsname\relax\def\natexlab#1{#1}\fi
\providecommand{\url}[1]{\href{#1}{#1}}
\providecommand{\dodoi}[1]{doi:~\href{http://doi.org/#1}{\nolinkurl{#1}}}
\providecommand{\doeprint}[1]{\href{http://ascl.net/#1}{\nolinkurl{http://ascl.net/#1}}}
\providecommand{\doarXiv}[1]{\href{https://arxiv.org/abs/#1}{\nolinkurl{https://arxiv.org/abs/#1}}}

\bibitem[{{Akiba} {et~al.}(2019){Akiba}, {Sano}, {Yanase}, {Ohta}, \&
  {Koyama}}]{Akiba2019}
{Akiba}, T., {Sano}, S., {Yanase}, T., {Ohta}, T., \& {Koyama}, M. 2019, arXiv
  e-prints, arXiv:1907.10902, \dodoi{10.48550/arXiv.1907.10902}

\bibitem[{{Alsing} {et~al.}(2019){Alsing}, {Charnock}, {Feeney}, \&
  {Wandelt}}]{Alsing2019}
{Alsing}, J., {Charnock}, T., {Feeney}, S., \& {Wandelt}, B. 2019, \mnras, 488,
  4440, \dodoi{10.1093/mnras/stz1960}

\bibitem[{{Behroozi} {et~al.}(2013){Behroozi}, {Wechsler}, \& {Wu}}]{rockstar}
{Behroozi}, P.~S., {Wechsler}, R.~H., \& {Wu}, H.-Y. 2013, \apj, 762, 109,
  \dodoi{10.1088/0004-637X/762/2/109}

\bibitem[{Bernardeau {et~al.}(2002)Bernardeau, Colombi, Gaztanaga, \&
  Scoccimarro}]{Bernardeau2002}
Bernardeau, F., Colombi, S., Gaztanaga, E., \& Scoccimarro, R. 2002, Physics
  Reports, 367, 1, \dodoi{10.1016/S0370-1573(02)00135-7}

\bibitem[{Beutler {et~al.}(2017)Beutler, Seo, Saito, Chuang, Cuesta,
  Eisenstein, {Gil-Mar{\'i}n}, Grieb, Hand, Kitaura, Modi, Nichol, Olmstead,
  Percival, Prada, S{\'a}nchez, {Rodriguez-Torres}, Ross, Ross, Schneider,
  Tinker, Tojeiro, \& {Vargas-Maga{\~n}a}}]{Beutler2017}
Beutler, F., Seo, H.-J., Saito, S., {et~al.} 2017, Monthly Notices of the Royal
  Astronomical Society, 466, 2242, \dodoi{10.1093/mnras/stw3298}

\bibitem[{{Bianchi} {et~al.}(2018){Bianchi}, {Burden}, {Percival}, {Brooks},
  {Cahn}, {Forero-Romero}, {Levi}, {Ross}, \& {Tarle}}]{Bianchi2018}
{Bianchi}, D., {Burden}, A., {Percival}, W.~J., {et~al.} 2018, \mnras, 481,
  2338, \dodoi{10.1093/mnras/sty2377}

\bibitem[{{Chen} {et~al.}(2022){Chen}, {Vlah}, \& {White}}]{Chen2022}
{Chen}, S.-F., {Vlah}, Z., \& {White}, M. 2022, \jcap, 2022, 008,
  \dodoi{10.1088/1475-7516/2022/02/008}

\bibitem[{Collaboration {et~al.}(2016{\natexlab{a}})Collaboration, Aghamousa,
  Aguilar, Ahlen, Alam, Allen, Prieto, Annis, Bailey, Balland, Ballester,
  Baltay, Beaufore, Bebek, Beers, Bell, Bernal, Besuner, Beutler, Blake,
  Bleuler, Blomqvist, Blum, Bolton, Briceno, Brooks, Brownstein,
  {Buckley-Geer}, Burden, Burtin, Busca, Cahn, Cai, {Cardiel-Sas}, Carlberg,
  Carton, Casas, Castander, {Cervantes-Cota}, Claybaugh, Close, Coker, Cole,
  Comparat, Cooper, Cousinou, Crocce, Cuby, Cunningham, Davis, Dawson, {de la
  Macorra}, De~Vicente, Delubac, Derwent, Dey, Dhungana, Ding, Doel, Duan,
  Ealet, Edelstein, Eftekharzadeh, Eisenstein, Elliott, Escoffier, Evatt,
  Fagrelius, Fan, Fanning, Farahi, Farihi, Favole, Feng, Fernandez, Findlay,
  Finkbeiner, Fitzpatrick, Flaugher, Flender, {Font-Ribera}, {Forero-Romero},
  Fosalba, Frenk, Fumagalli, Gaensicke, Gallo, {Garcia-Bellido}, Gaztanaga,
  Fusillo, Gerard, Gershkovich, Giannantonio, Gillet, {Gonzalez-de-Rivera},
  {Gonzalez-Perez}, Gott, Graur, Gutierrez, Guy, Habib, Heetderks, Heetderks,
  Heitmann, Hellwing, Herrera, Ho, Holland, Honscheid, Huff, Hutchinson,
  Huterer, Hwang, Laguna, Ishikawa, Jacobs, Jeffrey, Jelinsky, Jennings, Jiang,
  Jimenez, Johnson, Joyce, Jullo, Juneau, Kama, Karcher, Karkar, Kehoe,
  Kennamer, Kent, Kilbinger, Kim, Kirkby, Kisner, Kitanidis, Kneib, Koposov,
  Kovacs, Koyama, Kremin, Kron, Kronig, {Kueter-Young}, Lacey, Lafever, Lahav,
  Lambert, Lampton, Landriau, Lang, Lauer, Goff, Guillou, Van~Suu, Lee, Lee,
  Leitner, Lesser, Levi, L'Huillier, Li, Liang, Lin, Linder, Loebman,
  Luki{\'c}, Ma, MacCrann, Magneville, Makarem, Manera, Manser, Marshall,
  Martini, Massey, Matheson, McCauley, McDonald, McGreer, Meisner, Metcalfe,
  Miller, Miquel, Moustakas, Myers, Naik, Newman, Nichol, Nicola, {da Costa},
  Nie, Niz, Norberg, Nord, Norman, Nugent, O'Brien, Oh, Olsen, Padilla,
  Padmanabhan, Padmanabhan, {Palanque-Delabrouille}, Palmese, Pappalardo,
  P{\^a}ris, Park, Patej, Peacock, Peiris, Peng, Percival, Perruchot, Pieri,
  Pogge, Pollack, Poppett, Prada, Prakash, Probst, Rabinowitz, Raichoor, Ree,
  Refregier, Regal, Reid, Reil, Rezaie, Rockosi, Roe, Ronayette, Roodman, Ross,
  Ross, Rossi, Rozo, {Ruhlmann-Kleider}, Rykoff, Sabiu, Samushia, Sanchez,
  Sanchez, Schlegel, Schneider, Schubnell, Secroun, Seljak, Seo, Serrano,
  Shafieloo, Shan, Sharples, Sholl, Shourt, Silber, Silva, Sirk, Slosar, Smith,
  Smoot, Som, Song, Sprayberry, Staten, Stefanik, Tarle, Tie, Tinker, Tojeiro,
  Valdes, Valenzuela, Valluri, {Vargas-Magana}, Verde, Walker, Wang, Wang,
  Weaver, Weaverdyck, Wechsler, Weinberg, White, Yang, Yeche, Zhang, Zhao,
  Zheng, Zhou, Zhou, Zhu, Zou, \& Zu}]{desi1}
Collaboration, D., Aghamousa, A., Aguilar, J., {et~al.} 2016{\natexlab{a}},
  arXiv:1611.00036 [astro-ph].
\newblock \doarXiv{1611.00036}

\bibitem[{Collaboration {et~al.}(2016{\natexlab{b}})Collaboration, Aghamousa,
  Aguilar, Ahlen, Alam, Allen, Prieto, Annis, Bailey, Balland, Ballester,
  Baltay, Beaufore, Bebek, Beers, Bell, Bernal, Besuner, Beutler, Blake,
  Bleuler, Blomqvist, Blum, Bolton, Briceno, Brooks, Brownstein,
  {Buckley-Geer}, Burden, Burtin, Busca, Cahn, Cai, {Cardiel-Sas}, Carlberg,
  Carton, Casas, Castander, {Cervantes-Cota}, Claybaugh, Close, Coker, Cole,
  Comparat, Cooper, Cousinou, Crocce, Cuby, Cunningham, Davis, Dawson, {de la
  Macorra}, De~Vicente, Delubac, Derwent, Dey, Dhungana, Ding, Doel, Duan,
  Ealet, Edelstein, Eftekharzadeh, Eisenstein, Elliott, Escoffier, Evatt,
  Fagrelius, Fan, Fanning, Farahi, Farihi, Favole, Feng, Fernandez, Findlay,
  Finkbeiner, Fitzpatrick, Flaugher, Flender, {Font-Ribera}, {Forero-Romero},
  Fosalba, Frenk, Fumagalli, Gaensicke, Gallo, {Garcia-Bellido}, Gaztanaga,
  Fusillo, Gerard, Gershkovich, Giannantonio, Gillet, {Gonzalez-de-Rivera},
  {Gonzalez-Perez}, Gott, Graur, Gutierrez, Guy, Habib, Heetderks, Heetderks,
  Heitmann, Hellwing, Herrera, Ho, Holland, Honscheid, Huff, Hutchinson,
  Huterer, Hwang, Laguna, Ishikawa, Jacobs, Jeffrey, Jelinsky, Jennings, Jiang,
  Jimenez, Johnson, Joyce, Jullo, Juneau, Kama, Karcher, Karkar, Kehoe,
  Kennamer, Kent, Kilbinger, Kim, Kirkby, Kisner, Kitanidis, Kneib, Koposov,
  Kovacs, Koyama, Kremin, Kron, Kronig, {Kueter-Young}, Lacey, Lafever, Lahav,
  Lambert, Lampton, Landriau, Lang, Lauer, Goff, Guillou, Van~Suu, Lee, Lee,
  Leitner, Lesser, Levi, L'Huillier, Li, Liang, Lin, Linder, Loebman,
  Luki{\'c}, Ma, MacCrann, Magneville, Makarem, Manera, Manser, Marshall,
  Martini, Massey, Matheson, McCauley, McDonald, McGreer, Meisner, Metcalfe,
  Miller, Miquel, Moustakas, Myers, Naik, Newman, Nichol, Nicola, {da Costa},
  Nie, Niz, Norberg, Nord, Norman, Nugent, O'Brien, Oh, Olsen, Padilla,
  Padmanabhan, Padmanabhan, {Palanque-Delabrouille}, Palmese, Pappalardo,
  P{\^a}ris, Park, Patej, Peacock, Peiris, Peng, Percival, Perruchot, Pieri,
  Pogge, Pollack, Poppett, Prada, Prakash, Probst, Rabinowitz, Raichoor, Ree,
  Refregier, Regal, Reid, Reil, Rezaie, Rockosi, Roe, Ronayette, Roodman, Ross,
  Ross, Rossi, Rozo, {Ruhlmann-Kleider}, Rykoff, Sabiu, Samushia, Sanchez,
  Sanchez, Schlegel, Schneider, Schubnell, Secroun, Seljak, Seo, Serrano,
  Shafieloo, Shan, Sharples, Sholl, Shourt, Silber, Silva, Sirk, Slosar, Smith,
  Smoot, Som, Song, Sprayberry, Staten, Stefanik, Tarle, Tie, Tinker, Tojeiro,
  Valdes, Valenzuela, Valluri, {Vargas-Magana}, Verde, Walker, Wang, Wang,
  Weaver, Weaverdyck, Wechsler, Weinberg, White, Yang, Yeche, Zhang, Zhao,
  Zheng, Zhou, Zhou, Zhu, Zou, \& Zu}]{desi2}
---. 2016{\natexlab{b}}, arXiv:1611.00037 [astro-ph].
\newblock \doarXiv{1611.00037}

\bibitem[{{Contarini} {et~al.}(2023){Contarini}, {Pisani}, {Hamaus}, {Marulli},
  {Moscardini}, \& {Baldi}}]{Contarini2023}
{Contarini}, S., {Pisani}, A., {Hamaus}, N., {et~al.} 2023, \apj, 953, 46,
  \dodoi{10.3847/1538-4357/acde54}

\bibitem[{{Cranmer} {et~al.}(2020){Cranmer}, {Brehmer}, \&
  {Louppe}}]{Cranmer2020}
{Cranmer}, K., {Brehmer}, J., \& {Louppe}, G. 2020, Proceedings of the National
  Academy of Science, 117, 30055, \dodoi{10.1073/pnas.1912789117}

\bibitem[{{D'Amico} {et~al.}(2022){D'Amico}, {Donath}, {Lewandowski},
  {Senatore}, \& {Zhang}}]{DAmico2022}
{D'Amico}, G., {Donath}, Y., {Lewandowski}, M., {Senatore}, L., \& {Zhang}, P.
  2022, arXiv e-prints, arXiv:2206.08327, \dodoi{10.48550/arXiv.2206.08327}

\bibitem[{{d'Amico} {et~al.}(2020){d'Amico}, {Gleyzes}, {Kokron}, {Markovic},
  {Senatore}, {Zhang}, {Beutler}, \& {Gil-Mar{\'\i}n}}]{damico2020}
{d'Amico}, G., {Gleyzes}, J., {Kokron}, N., {et~al.} 2020, \jcap, 2020, 005,
  \dodoi{10.1088/1475-7516/2020/05/005}

\bibitem[{{Davis} {et~al.}(1985){Davis}, {Efstathiou}, {Frenk}, \&
  {White}}]{Davis1985}
{Davis}, M., {Efstathiou}, G., {Frenk}, C.~S., \& {White}, S.~D.~M. 1985, \apj,
  292, 371, \dodoi{10.1086/163168}

\bibitem[{Dawson {et~al.}(2013)Dawson, Schlegel, Ahn, Anderson, Aubourg,
  Bailey, Barkhouser, Bautista, Beifiori, Berlind, Bhardwaj, Bizyaev, Blake,
  Blanton, Blomqvist, Bolton, Borde, Bovy, Brandt, Brewington, Brinkmann,
  Brown, Brownstein, Bundy, Busca, Carithers, Carnero, Carr, Chen, Comparat,
  Connolly, Cope, Croft, Cuesta, {da Costa}, Davenport, Delubac, {de Putter},
  Dhital, Ealet, Ebelke, Eisenstein, Escoffier, Fan, Filiz~Ak, Finley,
  {Font-Ribera}, {G{\'e}nova-Santos}, Gunn, Guo, Haggard, Hall, Hamilton,
  Harris, Harris, Ho, Hogg, Holder, Honscheid, Huehnerhoff, Jordan, Jordan,
  Kauffmann, Kazin, Kirkby, Klaene, Kneib, Le~Goff, Lee, Long, Loomis,
  Lundgren, Lupton, Maia, Makler, Malanushenko, Malanushenko, Mandelbaum,
  Manera, Maraston, Margala, Masters, McBride, McDonald, McGreer, McMahon,
  Mena, {Miralda-Escud{\'e}}, {Montero-Dorta}, Montesano, Muna, Myers, Naugle,
  Nichol, Noterdaeme, Nuza, Olmstead, Oravetz, Oravetz, Owen, Padmanabhan,
  {Palanque-Delabrouille}, Pan, Parejko, P{\^a}ris, Percival,
  {P{\'e}rez-Fournon}, {P{\'e}rez-R{\`a}fols}, Petitjean, Pfaffenberger, Pforr,
  Pieri, Prada, {Price-Whelan}, Raddick, Rebolo, Rich, Richards, Rockosi, Roe,
  Ross, Ross, Rossi, {Rubi{\~n}o-Martin}, Samushia, S{\'a}nchez, Sayres,
  Schmidt, Schneider, Sc{\'o}ccola, Seo, Shelden, Sheldon, Shen, Shu, Slosar,
  Smee, Snedden, Stauffer, Steele, Strauss, Streblyanska, Suzuki, Swanson, Tal,
  Tanaka, Thomas, Tinker, Tojeiro, Tremonti, Vargas~Maga{\~n}a, Verde, Viel,
  Wake, Watson, Weaver, Weinberg, Weiner, West, White, {Wood-Vasey}, Yeche,
  Zehavi, Zhao, \& Zheng}]{Dawson2013}
Dawson, K.~S., Schlegel, D.~J., Ahn, C.~P., {et~al.} 2013, The Astronomical
  Journal, 145, 10, \dodoi{10.1088/0004-6256/145/1/10}

\bibitem[{Desjacques {et~al.}(2016)Desjacques, Jeong, \&
  Schmidt}]{Desjacques2016}
Desjacques, V., Jeong, D., \& Schmidt, F. 2016, arXiv:1611.09787 [astro-ph,
  physics:gr-qc, physics:hep-ph].
\newblock \doarXiv{1611.09787}

\bibitem[{{Eickenberg} {et~al.}(2022){Eickenberg}, {Allys}, {Moradinezhad
  Dizgah}, {Lemos}, {Massara}, {Abidi}, {Hahn}, {Hassan}, {Regaldo-Saint
  Blancard}, {Ho}, {Mallat}, {And{\'e}n}, \&
  {Villaescusa-Navarro}}]{Eickenberg2022}
{Eickenberg}, M., {Allys}, E., {Moradinezhad Dizgah}, A., {et~al.} 2022, arXiv
  e-prints, arXiv:2204.07646, \dodoi{10.48550/arXiv.2204.07646}

\bibitem[{Eisenstein {et~al.}(2011)Eisenstein, Weinberg, Agol, Aihara,
  Allende~Prieto, Anderson, Arns, Aubourg, Bailey, Balbinot, Barkhouser, Beers,
  Berlind, Bickerton, Bizyaev, Blanton, Bochanski, Bolton, Bosman, Bovy,
  Brandt, Breslauer, Brewington, Brinkmann, Brown, Brownstein, Burger, Busca,
  Campbell, Cargile, Carithers, Carlberg, Carr, Chang, Chen, Chiappini,
  Comparat, Connolly, Cortes, Croft, Cunha, {da Costa}, Davenport, Dawson,
  De~Lee, {Porto de Mello}, {de Simoni}, Dean, Dhital, Ealet, Ebelke,
  Edmondson, Eiting, Escoffier, Esposito, Evans, Fan, Femen{\'i}a~Castell{\'a},
  Dutra~Ferreira, Fitzgerald, Fleming, {Font-Ribera}, Ford, Frinchaboy,
  Garc{\'i}a~P{\'e}rez, Gaudi, Ge, Ghezzi, Gillespie, Gilmore, Girardi, Gott,
  Gould, Grebel, Gunn, Hamilton, Harding, Harris, Hawley, Hearty, Hennawi,
  Gonz{\'a}lez~Hern{\'a}ndez, Ho, Hogg, Holtzman, Honscheid, Inada, Ivans,
  Jiang, Jiang, Johnson, Jordan, Jordan, Kauffmann, Kazin, Kirkby, Klaene,
  Knapp, Kneib, Kochanek, Koesterke, Kollmeier, Kron, Lampeitl, Lang, Lawler,
  Le~Goff, Lee, Lee, Leisenring, Lin, Liu, Long, Loomis, Lucatello, Lundgren,
  Lupton, Ma, Ma, MacDonald, Mack, Mahadevan, Maia, Majewski, Makler,
  Malanushenko, Malanushenko, Mandelbaum, Maraston, Margala, Maseman, Masters,
  McBride, McDonald, McGreer, McMahon, Mena~Requejo, M{\'e}nard,
  {Miralda-Escud{\'e}}, Morrison, Mullally, Muna, Murayama, Myers, Naugle,
  Neto, Nguyen, Nichol, Nidever, O'Connell, Ogando, Olmstead, Oravetz,
  Padmanabhan, Paegert, {Palanque-Delabrouille}, Pan, Pandey, Parejko,
  P{\^a}ris, Pellegrini, Pepper, Percival, Petitjean, Pfaffenberger, Pforr,
  Phleps, Pichon, Pieri, Prada, {Price-Whelan}, Raddick, Ramos, Reid, Reyle,
  Rich, Richards, Rieke, Rieke, Rix, Robin, {Rocha-Pinto}, Rockosi, Roe,
  Rollinde, Ross, Ross, Rossetto, S{\'a}nchez, Santiago, Sayres, Schiavon,
  Schlegel, Schlesinger, Schmidt, Schneider, Sellgren, Shelden, Sheldon,
  Shetrone, Shu, Silverman, Simmerer, Simmons, Sivarani, Skrutskie, Slosar,
  Smee, Smith, Snedden, Stassun, Steele, Steinmetz, Stockett, Stollberg,
  Strauss, Szalay, Tanaka, Thakar, Thomas, Tinker, Tofflemire, Tojeiro,
  Tremonti, Vargas~Maga{\~n}a, Verde, Vogt, Wake, Wan, Wang, Weaver, White,
  White, Wilson, Wisniewski, {Wood-Vasey}, Yanny, Yasuda, Y{\`e}che, York,
  Young, Zasowski, Zehavi, \& Zhao}]{Eisenstein2011}
Eisenstein, D.~J., Weinberg, D.~H., Agol, E., {et~al.} 2011, The Astronomical
  Journal, 142, 72, \dodoi{10.1088/0004-6256/142/3/72}

\bibitem[{{Guo} {et~al.}(2012){Guo}, {Zehavi}, \& {Zheng}}]{Guo2012}
{Guo}, H., {Zehavi}, I., \& {Zheng}, Z. 2012, \apj, 756, 127,
  \dodoi{10.1088/0004-637X/756/2/127}

\bibitem[{{Hadzhiyska} {et~al.}(2022){Hadzhiyska}, {Eisenstein}, {Bose},
  {Garrison}, \& {Maksimova}}]{compaso}
{Hadzhiyska}, B., {Eisenstein}, D., {Bose}, S., {Garrison}, L.~H., \&
  {Maksimova}, N. 2022, \mnras, 509, 501, \dodoi{10.1093/mnras/stab2980}

\bibitem[{Hadzhiyska {et~al.}(2021)Hadzhiyska, Liu, Somerville, Gabrielpillai,
  Bose, Eisenstein, \& Hernquist}]{hadzhiyska2021}
Hadzhiyska, B., Liu, S., Somerville, R.~S., {et~al.} 2021, Monthly Notices of
  the Royal Astronomical Society, 508, 698, \dodoi{10.1093/mnras/stab2564}

\bibitem[{{Hahn} {et~al.}(2017){Hahn}, {Scoccimarro}, {Blanton}, {Tinker}, \&
  {Rodr{\'\i}guez-Torres}}]{Hahn2017}
{Hahn}, C., {Scoccimarro}, R., {Blanton}, M.~R., {Tinker}, J.~L., \&
  {Rodr{\'\i}guez-Torres}, S.~A. 2017, \mnras, 467, 1940,
  \dodoi{10.1093/mnras/stx185}

\bibitem[{{Hahn} \& {Villaescusa-Navarro}(2021)}]{Hahn2021}
{Hahn}, C., \& {Villaescusa-Navarro}, F. 2021, \jcap, 2021, 029,
  \dodoi{10.1088/1475-7516/2021/04/029}

\bibitem[{{Hahn} {et~al.}(2022){Hahn}, {Eickenberg}, {Ho}, {Hou}, {Lemos},
  {Massara}, {Modi}, {Moradinezhad Dizgah}, {R{\'e}galdo-Saint Blancard}, \&
  {Abidi}}]{simbig_letter}
{Hahn}, C., {Eickenberg}, M., {Ho}, S., {et~al.} 2022, arXiv e-prints,
  arXiv:2211.00723, \dodoi{10.48550/arXiv.2211.00723}

\bibitem[{{Hahn} {et~al.}(2023{\natexlab{a}}){Hahn}, {Lemos}, {Parker},
  {R{\'e}galdo-Saint Blancard}, {Eickenberg}, {Ho}, {Hou}, {Massara}, {Modi},
  {Moradinezhad Dizgah}, \& {Spergel}}]{simbig_wave2}
{Hahn}, C., {Lemos}, P., {Parker}, L., {et~al.} 2023{\natexlab{a}}, arXiv
  e-prints, arXiv:2310.15246, \dodoi{10.48550/arXiv.2310.15246}

\bibitem[{{Hahn} {et~al.}(2023{\natexlab{b}}){Hahn}, {Eickenberg}, {Ho}, {Hou},
  {Lemos}, {Massara}, {Modi}, {Moradinezhad Dizgah}, {R{\'e}galdo-Saint
  Blancard}, \& {Abidi}}]{simbig_challenge}
{Hahn}, C., {Eickenberg}, M., {Ho}, S., {et~al.} 2023{\natexlab{b}}, \jcap,
  2023, 010, \dodoi{10.1088/1475-7516/2023/04/010}

\bibitem[{{Hahn} {et~al.}(2023{\natexlab{c}}){Hahn}, {Eickenberg}, {Ho}, {Hou},
  {Lemos}, {Massara}, {Modi}, {Moradinezhad Dizgah}, {Parker}, \&
  {R{\'e}galdo-Saint Blancard}}]{simbig_bk}
---. 2023{\natexlab{c}}, arXiv e-prints, arXiv:2310.15243,
  \dodoi{10.48550/arXiv.2310.15243}

\bibitem[{{Hahn} {et~al.}(2023{\natexlab{d}}){Hahn}, {Wilson}, {Ruiz-Macias},
  {Cole}, {Weinberg}, {Moustakas}, {Kremin}, {Tinker}, {Smith}, {Wechsler},
  {Ahlen}, {Alam}, {Bailey}, {Brooks}, {Cooper}, {Davis}, {Dawson}, {Dey},
  {Dey}, {Eftekharzadeh}, {Eisenstein}, {Fanning}, {Forero-Romero}, {Frenk},
  {Gazta{\~n}aga}, {Gontcho A Gontcho}, {Guy}, {Honscheid}, {Ishak}, {Juneau},
  {Kehoe}, {Kisner}, {Lan}, {Landriau}, {Le Guillou}, {Levi}, {Magneville},
  {Martini}, {Meisner}, {Myers}, {Nie}, {Norberg}, {Palanque-Delabrouille},
  {Percival}, {Poppett}, {Prada}, {Raichoor}, {Ross}, {Safonova}, {Saulder},
  {Schlafly}, {Schlegel}, {Sierra-Porta}, {Tarle}, {Weaver}, {Y{\`e}che},
  {Zarrouk}, {Zhou}, {Zhou}, \& {Zou}}]{Hahn2023}
{Hahn}, C., {Wilson}, M.~J., {Ruiz-Macias}, O., {et~al.} 2023{\natexlab{d}},
  \aj, 165, 253, \dodoi{10.3847/1538-3881/accff8}

\bibitem[{Hand {et~al.}(2018)Hand, Feng, Beutler, Li, Modi, Seljak, \&
  Slepian}]{nbodykit}
Hand, N., Feng, Y., Beutler, F., {et~al.} 2018, Astron. J., 156, 160,
  \dodoi{10.3847/1538-3881/aadae0}

\bibitem[{{Hou} {et~al.}(2023){Hou}, {Moradinezhad Dizgah}, {Hahn}, \&
  {Massara}}]{Hou2023Fisher}
{Hou}, J., {Moradinezhad Dizgah}, A., {Hahn}, C., \& {Massara}, E. 2023, \jcap,
  2023, 045, \dodoi{10.1088/1475-7516/2023/03/045}

\bibitem[{{Hou} {et~al.}(2024){Hou}, {Moradinezhad Dizgah}, {Hahn},
  {Eickenberg}, {Ho}, {Lemos}, {Massara}, {Modi}, {Parker}, \&
  {R{\'e}galdo-Saint Blancard}}]{simbig_skewspec}
{Hou}, J., {Moradinezhad Dizgah}, A., {Hahn}, C., {et~al.} 2024, arXiv
  e-prints, arXiv:2401.15074, \dodoi{10.48550/arXiv.2401.15074}

\bibitem[{{Ivanov} {et~al.}(2023){Ivanov}, {Philcox}, {Cabass}, {Nishimichi},
  {Simonovi{\'c}}, \& {Zaldarriaga}}]{Ivanov2023}
{Ivanov}, M.~M., {Philcox}, O. H.~E., {Cabass}, G., {et~al.} 2023, \prd, 107,
  083515, \dodoi{10.1103/PhysRevD.107.083515}

\bibitem[{Ivanov {et~al.}(2020)Ivanov, Simonovi{\'c}, \&
  Zaldarriaga}]{Ivanov2020}
Ivanov, M.~M., Simonovi{\'c}, M., \& Zaldarriaga, M. 2020, Journal of Cosmology
  and Astroparticle Physics, 2020, 042, \dodoi{10.1088/1475-7516/2020/05/042}

\bibitem[{{Kingma} \& {Ba}(2014)}]{Kingma2014}
{Kingma}, D.~P., \& {Ba}, J. 2014, arXiv e-prints, arXiv:1412.6980,
  \dodoi{10.48550/arXiv.1412.6980}

\bibitem[{{Kobayashi} {et~al.}(2022){Kobayashi}, {Nishimichi}, {Takada}, \&
  {Miyatake}}]{Kobayashi2022}
{Kobayashi}, Y., {Nishimichi}, T., {Takada}, M., \& {Miyatake}, H. 2022, \prd,
  105, 083517, \dodoi{10.1103/PhysRevD.105.083517}

\bibitem[{{Kreisch} {et~al.}(2022){Kreisch}, {Pisani}, {Villaescusa-Navarro},
  {Spergel}, {Wandelt}, {Hamaus}, \& {Bayer}}]{Kreisch2022}
{Kreisch}, C.~D., {Pisani}, A., {Villaescusa-Navarro}, F., {et~al.} 2022, \apj,
  935, 100, \dodoi{10.3847/1538-4357/ac7d4b}

\bibitem[{{Lakshminarayanan} {et~al.}(2016){Lakshminarayanan}, {Pritzel}, \&
  {Blundell}}]{Lakshminarayanan2016}
{Lakshminarayanan}, B., {Pritzel}, A., \& {Blundell}, C. 2016, arXiv e-prints,
  arXiv:1612.01474, \dodoi{10.48550/arXiv.1612.01474}

\bibitem[{{Laureijs} {et~al.}(2011){Laureijs}, {Amiaux}, {Arduini},
  {Augu{\`e}res}, {Brinchmann}, {Cole}, {Cropper}, {Dabin}, {Duvet}, {Ealet},
  {Garilli}, {Gondoin}, {Guzzo}, {Hoar}, {Hoekstra}, {Holmes}, {Kitching},
  {Maciaszek}, {Mellier}, {Pasian}, {Percival}, {Rhodes}, {Saavedra Criado},
  {Sauvage}, {Scaramella}, {Valenziano}, {Warren}, {Bender}, {Castander},
  {Cimatti}, {Le F{\`e}vre}, {Kurki-Suonio}, {Levi}, {Lilje}, {Meylan},
  {Nichol}, {Pedersen}, {Popa}, {Rebolo Lopez}, {Rix}, {Rottgering},
  {Zeilinger}, {Grupp}, {Hudelot}, {Massey}, {Meneghetti}, {Miller}, {Paltani},
  {Paulin-Henriksson}, {Pires}, {Saxton}, {Schrabback}, {Seidel}, {Walsh},
  {Aghanim}, {Amendola}, {Bartlett}, {Baccigalupi}, {Beaulieu}, {Benabed},
  {Cuby}, {Elbaz}, {Fosalba}, {Gavazzi}, {Helmi}, {Hook}, {Irwin}, {Kneib},
  {Kunz}, {Mannucci}, {Moscardini}, {Tao}, {Teyssier}, {Weller}, {Zamorani},
  {Zapatero Osorio}, {Boulade}, {Foumond}, {Di Giorgio}, {Guttridge}, {James},
  {Kemp}, {Martignac}, {Spencer}, {Walton}, {Bl{\"u}mchen}, {Bonoli},
  {Bortoletto}, {Cerna}, {Corcione}, {Fabron}, {Jahnke}, {Ligori}, {Madrid},
  {Martin}, {Morgante}, {Pamplona}, {Prieto}, {Riva}, {Toledo}, {Trifoglio},
  {Zerbi}, {Abdalla}, {Douspis}, {Grenet}, {Borgani}, {Bouwens}, {Courbin},
  {Delouis}, {Dubath}, {Fontana}, {Frailis}, {Grazian}, {Koppenh{\"o}fer},
  {Mansutti}, {Melchior}, {Mignoli}, {Mohr}, {Neissner}, {Noddle}, {Poncet},
  {Scodeggio}, {Serrano}, {Shane}, {Starck}, {Surace}, {Taylor},
  {Verdoes-Kleijn}, {Vuerli}, {Williams}, {Zacchei}, {Altieri}, {Escudero
  Sanz}, {Kohley}, {Oosterbroek}, {Astier}, {Bacon}, {Bardelli}, {Baugh},
  {Bellagamba}, {Benoist}, {Bianchi}, {Biviano}, {Branchini}, {Carbone},
  {Cardone}, {Clements}, {Colombi}, {Conselice}, {Cresci}, {Deacon}, {Dunlop},
  {Fedeli}, {Fontanot}, {Franzetti}, {Giocoli}, {Garcia-Bellido}, {Gow},
  {Heavens}, {Hewett}, {Heymans}, {Holland}, {Huang}, {Ilbert}, {Joachimi},
  {Jennins}, {Kerins}, {Kiessling}, {Kirk}, {Kotak}, {Krause}, {Lahav}, {van
  Leeuwen}, {Lesgourgues}, {Lombardi}, {Magliocchetti}, {Maguire}, {Majerotto},
  {Maoli}, {Marulli}, {Maurogordato}, {McCracken}, {McLure}, {Melchiorri},
  {Merson}, {Moresco}, {Nonino}, {Norberg}, {Peacock}, {Pello}, {Penny},
  {Pettorino}, {Di Porto}, {Pozzetti}, {Quercellini}, {Radovich}, {Rassat},
  {Roche}, {Ronayette}, {Rossetti}, {Sartoris}, {Schneider}, {Semboloni},
  {Serjeant}, {Simpson}, {Skordis}, {Smadja}, {Smartt}, {Spano}, {Spiro},
  {Sullivan}, {Tilquin}, {Trotta}, {Verde}, {Wang}, {Williger}, {Zhao},
  {Zoubian}, \& {Zucca}}]{Laureijs2011}
{Laureijs}, R., {Amiaux}, J., {Arduini}, S., {et~al.} 2011, arXiv e-prints,
  arXiv:1110.3193, \dodoi{10.48550/arXiv.1110.3193}

\bibitem[{{Lemos} {et~al.}(2023){Lemos}, {Parker}, {Hahn}, {Ho}, {Eickenberg},
  {Hou}, {Massara}, {Modi}, {Moradinezhad Dizgah}, {R{\'e}galdo-Saint
  Blancard}, \& {Spergel}}]{simbig_cnn}
{Lemos}, P., {Parker}, L.~H., {Hahn}, C., {et~al.} 2023, in Machine Learning
  for Astrophysics, 18, \dodoi{10.48550/arXiv.2310.15256}

\bibitem[{{Maksimova} {et~al.}(2021){Maksimova}, {Garrison}, {Eisenstein},
  {Hadzhiyska}, {Bose}, \& {Satterthwaite}}]{abacus}
{Maksimova}, N.~A., {Garrison}, L.~H., {Eisenstein}, D.~J., {et~al.} 2021,
  \mnras, 508, 4017, \dodoi{10.1093/mnras/stab2484}

\bibitem[{{Massara} {et~al.}(2021){Massara}, {Villaescusa-Navarro}, {Ho},
  {Dalal}, \& {Spergel}}]{Massara2021}
{Massara}, E., {Villaescusa-Navarro}, F., {Ho}, S., {Dalal}, N., \& {Spergel},
  D.~N. 2021, \prl, 126, 011301, \dodoi{10.1103/PhysRevLett.126.011301}

\bibitem[{{Massara} {et~al.}(2023){Massara}, {Villaescusa-Navarro}, {Hahn},
  {Abidi}, {Eickenberg}, {Ho}, {Lemos}, {Dizgah}, \& {Blancard}}]{Massara2023}
{Massara}, E., {Villaescusa-Navarro}, F., {Hahn}, C., {et~al.} 2023, \apj, 951,
  70, \dodoi{10.3847/1538-4357/acd44d}

\bibitem[{{Naidoo} {et~al.}(2022){Naidoo}, {Massara}, \& {Lahav}}]{Naidoo2022}
{Naidoo}, K., {Massara}, E., \& {Lahav}, O. 2022, \mnras, 513, 3596,
  \dodoi{10.1093/mnras/stac1138}

\bibitem[{{Paillas} {et~al.}(2023{\natexlab{a}}){Paillas}, {Cuesta-Lazaro},
  {Zarrouk}, {Cai}, {Percival}, {Nadathur}, {Pinon}, {de Mattia}, \&
  {Beutler}}]{Paillas2023}
{Paillas}, E., {Cuesta-Lazaro}, C., {Zarrouk}, P., {et~al.} 2023{\natexlab{a}},
  \mnras, 522, 606, \dodoi{10.1093/mnras/stad1017}

\bibitem[{{Paillas} {et~al.}(2023{\natexlab{b}}){Paillas}, {Cuesta-Lazaro},
  {Percival}, {Nadathur}, {Cai}, {Yuan}, {Beutler}, {de Mattia}, {Eisenstein},
  {Forero-Sanchez}, {Padilla}, {Pinon}, {Ruhlmann-Kleider}, {S{\'a}nchez},
  {Valogiannis}, \& {Zarrouk}}]{Paillas2023CMASS}
{Paillas}, E., {Cuesta-Lazaro}, C., {Percival}, W.~J., {et~al.}
  2023{\natexlab{b}}, arXiv e-prints, arXiv:2309.16541,
  \dodoi{10.48550/arXiv.2309.16541}

\bibitem[{{Papamakarios} {et~al.}(2017){Papamakarios}, {Pavlakou}, \&
  {Murray}}]{Papamakarios2017}
{Papamakarios}, G., {Pavlakou}, T., \& {Murray}, I. 2017, arXiv e-prints,
  arXiv:1705.07057, \dodoi{10.48550/arXiv.1705.07057}

\bibitem[{{Philcox} {et~al.}(2021){Philcox}, {Aviles}, \&
  {Massara}}]{Philcox2021}
{Philcox}, O. H.~E., {Aviles}, A., \& {Massara}, E. 2021, \jcap, 2021, 038,
  \dodoi{10.1088/1475-7516/2021/03/038}

\bibitem[{{Philcox} {et~al.}(2020){Philcox}, {Massara}, \&
  {Spergel}}]{Philcox2020}
{Philcox}, O. H.~E., {Massara}, E., \& {Spergel}, D.~N. 2020, \prd, 102,
  043516, \dodoi{10.1103/PhysRevD.102.043516}

\bibitem[{{Pinol} {et~al.}(2017){Pinol}, {Cahn}, {Hand}, {Seljak}, \&
  {White}}]{Pinol2017}
{Pinol}, L., {Cahn}, R.~N., {Hand}, N., {Seljak}, U., \& {White}, M. 2017,
  \jcap, 2017, 008, \dodoi{10.1088/1475-7516/2017/04/008}

\bibitem[{{R{\'e}galdo-Saint Blancard} {et~al.}(2023){R{\'e}galdo-Saint
  Blancard}, {Hahn}, {Ho}, {Hou}, {Lemos}, {Massara}, {Modi}, {Moradinezhad
  Dizgah}, {Parker}, {Yao}, \& {Eickenberg}}]{simbig_WST}
{R{\'e}galdo-Saint Blancard}, B., {Hahn}, C., {Ho}, S., {et~al.} 2023, arXiv
  e-prints, arXiv:2310.15250, \dodoi{10.48550/arXiv.2310.15250}

\bibitem[{Sheth {et~al.}(2005)Sheth, Connolly, \& Skibba}]{Sheth2005}
Sheth, R.~K., Connolly, A.~J., \& Skibba, R. 2005, Submitted to: Mon. Not. Roy.
  Astron. Soc.
\newblock \doarXiv{astro-ph/0511773}

\bibitem[{Spergel {et~al.}(2015)Spergel, Gehrels, Baltay, Bennett,
  Breckinridge, Donahue, Dressler, Gaudi, Greene, Guyon, Hirata, Kalirai,
  Kasdin, Macintosh, Moos, Perlmutter, Postman, Rauscher, Rhodes, Wang,
  Weinberg, Benford, Hudson, Jeong, Mellier, Traub, Yamada, Capak, Colbert,
  Masters, Penny, Savransky, Stern, Zimmerman, Barry, Bartusek, Carpenter,
  Cheng, Content, Dekens, Demers, Grady, Jackson, Kuan, Kruk, Melton, Nemati,
  Parvin, Poberezhskiy, Peddie, Ruffa, Wallace, Whipple, Wollack, \&
  Zhao}]{Spergel2015}
Spergel, D., Gehrels, N., Baltay, C., {et~al.} 2015, Wide-{{Field InfrarRed
  Survey Telescope-Astrophysics Focused Telescope Assets WFIRST-AFTA}} 2015
  {{Report}}

\bibitem[{Takada {et~al.}(2014)Takada, Ellis, Chiba, Greene, Aihara, Arimoto,
  Bundy, Cohen, Dor{\'e}, Graves, Gunn, Heckman, Hirata, Ho, Kneib,
  Le~F{\`e}vre, Lin, More, Murayama, Nagao, Ouchi, Seiffert, Silverman,
  Sodr{\'e}, Spergel, Strauss, Sugai, Suto, Takami, \& Wyse}]{Takada2014}
Takada, M., Ellis, R.~S., Chiba, M., {et~al.} 2014, Publications of the
  Astronomical Society of Japan, 66, R1, \dodoi{10.1093/pasj/pst019}

\bibitem[{Talts {et~al.}(2020)Talts, Betancourt, Simpson, Vehtari, \&
  Gelman}]{Talts2020}
Talts, S., Betancourt, M., Simpson, D., Vehtari, A., \& Gelman, A. 2020,
  arXiv:1804.06788 [stat].
\newblock \doarXiv{1804.06788}

\bibitem[{Tamura {et~al.}(2016)Tamura, Takato, Shimono, Moritani, Yabe,
  Ishizuka, Ueda, Kamata, Aghazarian, Arnouts, Barban, Barkhouser, Borges,
  Braun, Carr, Chabaud, Chang, Chen, Chiba, Chou, Chu, Cohen, {de Almeida}, {de
  Oliveira}, {de Oliveira}, Dekany, Dohlen, {dos Santos}, {dos Santos}, Ellis,
  Fabricius, Ferrand, Ferreira, Golebiowski, Greene, Gross, Gunn, Hammond,
  Harding, Hart, Heckman, Hirata, Ho, Hope, Hovland, Hsu, Hu, Huang, Jaquet,
  Jing, Karr, Kimura, King, Komatsu, Le~Brun, Le~F{\`e}vre, Le~Fur, Le~Mignant,
  Ling, Loomis, Lupton, Madec, Mao, Marrara, {Mendes de Oliveira}, Minowa,
  Morantz, Murayama, Murray, Ohyama, Orndorff, Pascal, Pereira, Reiley,
  Reinecke, Ritter, Roberts, Schwochert, Seiffert, Smee, Sodre, Spergel,
  Steinkraus, Strauss, Surace, Suto, Suzuki, Swinbank, Tait, Takada, Tamura,
  Tanaka, Tresse, Verducci, Vibert, Vidal, Wang, Wen, Yan, \&
  Yasuda}]{Tamura2016}
Tamura, N., Takato, N., Shimono, A., {et~al.} 2016, in Ground-Based and
  {{Airborne Instrumentation}} for {{Astronomy VI}}, Vol. 9908, {eprint:
  arXiv:1608.01075}, 99081M, \dodoi{10.1117/12.2232103}

\bibitem[{Tejero-Cantero {et~al.}(2020)Tejero-Cantero, Boelts, Deistler,
  Lueckmann, Durkan, Gonçalves, Greenberg, \& Macke}]{sbi}
Tejero-Cantero, A., Boelts, J., Deistler, M., {et~al.} 2020, Journal of Open
  Source Software, 5, 2505, \dodoi{10.21105/joss.02505}

\bibitem[{Vakili \& Hahn(2019)}]{vakili2019}
Vakili, M., \& Hahn, C. 2019, The Astrophysical Journal, 872, 115,
  \dodoi{10.3847/1538-4357/aaf1a1}

\bibitem[{{Valogiannis} \& {Bean}(2018)}]{Valogiannis2018}
{Valogiannis}, G., \& {Bean}, R. 2018, \prd, 97, 023535,
  \dodoi{10.1103/PhysRevD.97.023535}

\bibitem[{{Valogiannis} \& {Dvorkin}(2022)}]{Valogiannis2022}
{Valogiannis}, G., \& {Dvorkin}, C. 2022, \prd, 105, 103534,
  \dodoi{10.1103/PhysRevD.105.103534}

\bibitem[{{Valogiannis} {et~al.}(2023){Valogiannis}, {Yuan}, \&
  {Dvorkin}}]{Valogiannis2023}
{Valogiannis}, G., {Yuan}, S., \& {Dvorkin}, C. 2023, arXiv e-prints,
  arXiv:2310.16116, \dodoi{10.48550/arXiv.2310.16116}

\bibitem[{{Villaescusa-Navarro} {et~al.}(2020){Villaescusa-Navarro}, {Hahn},
  {Massara}, {Banerjee}, {Delgado}, {Ramanah}, {Charnock}, {Giusarma}, {Li},
  {Allys}, {Brochard}, {Uhlemann}, {Chiang}, {He}, {Pisani}, {Obuljen}, {Feng},
  {Castorina}, {Contardo}, {Kreisch}, {Nicola}, {Alsing}, {Scoccimarro},
  {Verde}, {Viel}, {Ho}, {Mallat}, {Wandelt}, \& {Spergel}}]{quijote}
{Villaescusa-Navarro}, F., {Hahn}, C., {Massara}, E., {et~al.} 2020, \apjs,
  250, 2, \dodoi{10.3847/1538-4365/ab9d82}

\bibitem[{{Wang} {et~al.}(2022){Wang}, {Zhai}, {Alavi}, {Massara}, {Pisani},
  {Benson}, {Hirata}, {Samushia}, {Weinberg}, {Colbert}, {Dor{\'e}}, {Eifler},
  {Heinrich}, {Ho}, {Krause}, {Padmanabhan}, {Spergel}, \&
  {Teplitz}}]{Wang2022}
{Wang}, Y., {Zhai}, Z., {Alavi}, A., {et~al.} 2022, \apj, 928, 1,
  \dodoi{10.3847/1538-4357/ac4973}

\bibitem[{White(2016)}]{White2016}
White, M. 2016, JCAP, 1611, 057, \dodoi{10.1088/1475-7516/2016/11/057}

\bibitem[{{White} \& {Padmanabhan}(2009)}]{White2009}
{White}, M., \& {Padmanabhan}, N. 2009, \mnras, 395, 2381,
  \dodoi{10.1111/j.1365-2966.2009.14732.x}

\bibitem[{{Yao} {et~al.}(2023){Yao}, {R{\'e}galdo-Saint Blancard}, \&
  {Domke}}]{Yao2023}
{Yao}, Y., {R{\'e}galdo-Saint Blancard}, B., \& {Domke}, J. 2023, arXiv
  e-prints, arXiv:2310.17009, \dodoi{10.48550/arXiv.2310.17009}

\bibitem[{Zentner {et~al.}(2016)Zentner, Hearin, van~den Bosch, Lange, \&
  Villarreal}]{zentner2016}
Zentner, A.~R., Hearin, A., van~den Bosch, F.~C., Lange, J.~U., \& Villarreal,
  A. 2016, arXiv:1606.07817 [astro-ph].
\newblock \doarXiv{1606.07817}

\bibitem[{{Zhai} {et~al.}(2023){Zhai}, {Tinker}, {Banerjee}, {DeRose}, {Guo},
  {Mao}, {McLaughlin}, {Storey-Fisher}, \& {Wechsler}}]{Zhai2023Aemulus}
{Zhai}, Z., {Tinker}, J.~L., {Banerjee}, A., {et~al.} 2023, \apj, 948, 99,
  \dodoi{10.3847/1538-4357/acc65b}

\bibitem[{{Zheng} {et~al.}(2007){Zheng}, {Coil}, \& {Zehavi}}]{Zheng2007}
{Zheng}, Z., {Coil}, A.~L., \& {Zehavi}, I. 2007, \apj, 667, 760,
  \dodoi{10.1086/521074}

\end{thebibliography}
\bibliographystyle{aasjournal}

\end{document}